\documentclass[conference]{IEEEtran}

\usepackage{cite}                 
\usepackage{graphicx}
\usepackage{amsmath,amssymb,amsfonts}
\usepackage{booktabs}             
\usepackage{amssymb}
\usepackage{multirow}             
\usepackage{array}
\usepackage{xcolor}
\usepackage{textcomp}

\usepackage{subcaption}
\usepackage{algorithm}
\usepackage{algpseudocode}

\usepackage[hidelinks]{hyperref}

\begin{document}

\title{EdgeXpert: An Edge Device for \\ Memory-Efficient LLM Inference with \\
Mixture-of-Experts and Speculative Decoding}

\author{
\IEEEauthorblockN{Sangwoo Ha}
\IEEEauthorblockA{\textit{KAIST} \\
Daejeon, South Korea \\
sangwoo\_ha@kaist.ac.kr}
\and
\IEEEauthorblockN{Hyunwoo Seo}
\IEEEauthorblockA{\textit{KAIST} \\
Daejeon, South Korea \\
shw4166@kaist.ac.kr}
\and
\IEEEauthorblockN{Yurim Jo}
\IEEEauthorblockA{\textit{KAIST} \\
Daejeon, South Korea \\
yurim.jo@kaist.ac.kr}
\and
\IEEEauthorblockN{Youngjin Moon}
\IEEEauthorblockA{\textit{KAIST} \\
Daejeon, South Korea \\
yj.moon@kaist.ac.kr}
\and
\IEEEauthorblockN{Hoi-Jun Yoo}
\IEEEauthorblockA{\textit{KAIST} \\
Daejeon, South Korea \\
hjyoo@kaist.ac.kr}
}

\maketitle

\begin{abstract}
On-device deployment of Large Language Models (LLMs) has become essential for personalized edge applications. A primary bottleneck is external memory access (EMA) in feed-forward network (FFN) layers. Speculative decoding and mixture-of-experts (MoE) are promising solutions. Speculative decoding reduces the number of decoding stages by generating multiple tokens per stage, and MoE minimizes per-stage cost through sparse expert activation. However, there is an incompatibility when combining these two techniques.

We propose EdgeXpert, a software-hardware co-designed LLM accelerator that resolves this incompatibility. In the prefill stage, the prompt-wise expert reuse reformulates routing as prompt-level expert reuse rather than independent per-token expert selection. It identifies important tokens using a lightweight encoder, constructs a shared expert set from them, and routes less important tokens with a reduced expert budget to lower expert EMA. In the decode stage, depth-aware expert coalescing exploits the contextual similarity and mutual exclusivity of same-depth candidate tokens. Rather than loading the union of all required channels, EdgeXpert loads only salient channels and applies computational calibration to recover accuracy without additional memory access. Synthesized in Samsung 28nm technology at 800 MHz, EdgeXpert achieves up to 56.3\% latency reduction and 44.1\% energy reduction compared to prior works \cite{moe-pruner, edgemoe, smolpu}, while maintaining near-baseline accuracy.

\end{abstract}

\begin{IEEEkeywords}
Edge device, large language model, mixture-of-experts, speculative decoding
\end{IEEEkeywords}


\section{Introduction}
\label{sec:intro}

\begin{figure}[t]
  \centering
    \hspace*{\fill} 
    \includegraphics[width=0.95\linewidth]{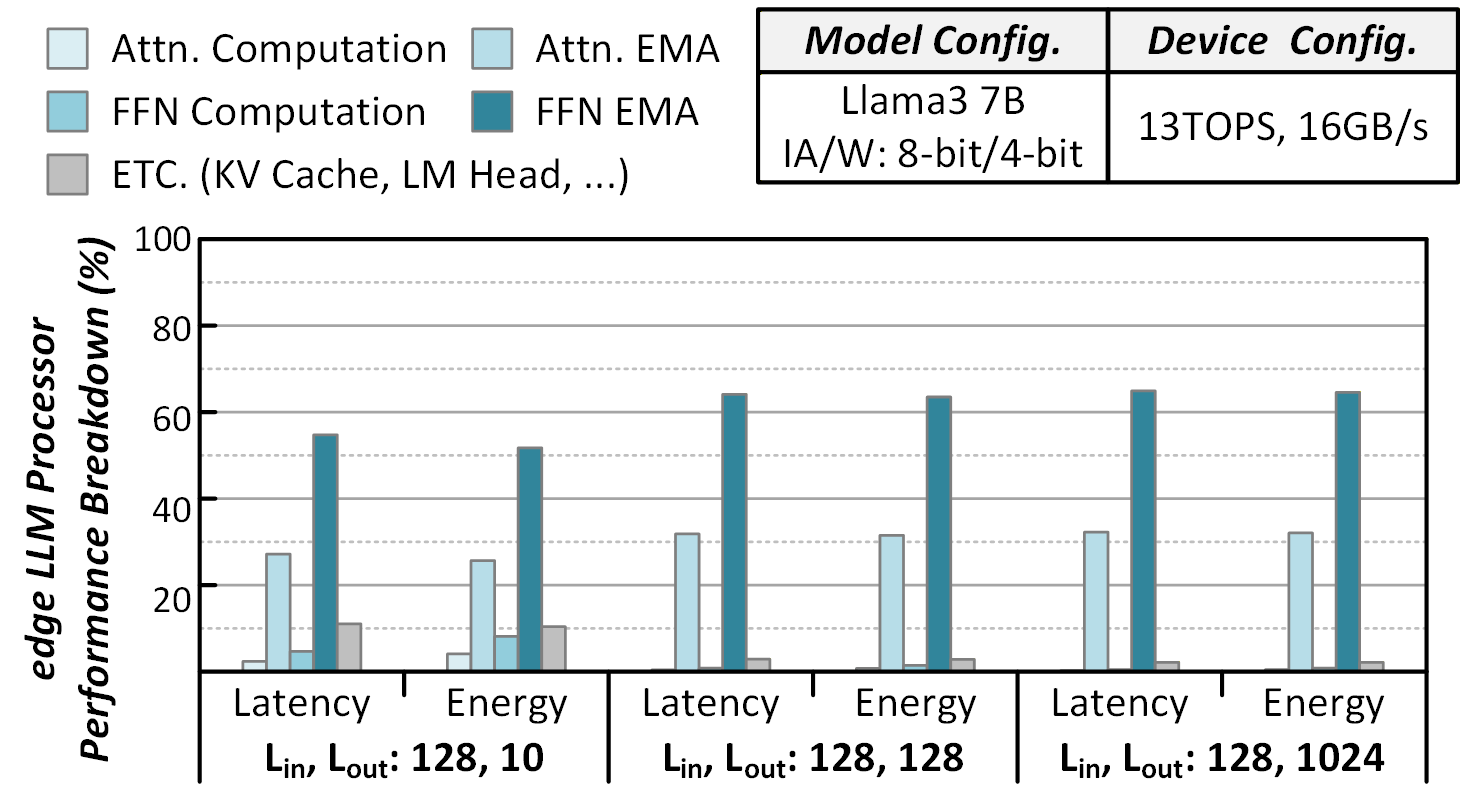}
    \hspace*{\fill} 
  \caption{Overall LLM system breakdown in edge devices.}
  \label{fig_1}
\end{figure}

Large language models (LLMs) \cite{gpt, llama, gemini} are being adopted across a wide range of applications, and there is growing interest in deploying them directly on edge devices for personalized, low-latency experiences. However, on-device inference is constrained by memory bandwidth and capacity. Edge LLMs typically operate in single-batch, autoregressive inference, which requires all model parameters to be loaded from external memory for each output token. This characteristic incurs substantial external memory access (EMA) latency and energy, becoming a primary cause of system bottlenecks. As shown in Figure~\ref{fig_1}, the EMA of the weight parameters constitutes the largest overhead. Specifically, the weight EMA in the feed-forward network (FFN) layers is the most dominant factor, and this overhead grows linearly with the number of decoding stages. While recent works have focused on mitigating KV cache overhead in long-context scenarios \cite{squeezed_attention, alisa}, on-device contexts are typically short (tens to a thousand tokens) \cite{broca, asplos_ondevicellm}, making KV cache a less significant bottleneck. Consequently, weight EMA emerges as the primary overhead for on-device LLM. For a conventional autoregressive decoding LLM, the weight EMA per token is defined by the total number of weight parameters ($\mathbf{P_{\text{total}}}$) that must be accessed from external memory per output token, as expressed in Equation~\eqref{eq:ema_conv}.  Previous works \cite{c-transformer, mecla, broca} have focused on reducing $\mathbf{P_{\text{total}}}$ by employing various model compression techniques, such as quantization, pruning, and tensor decomposition.

\vspace{-3mm}
\begin{equation}
\label{eq:ema_conv}
EMA_{\text{W, Conventional}} = P_{\text{total}}
\end{equation}
%

The substantial computation overhead associated with FFN layers has driven the exploration of more efficient models. The mixture-of-experts (MoE) models \cite{granite, qwen3, deepseekmoe, glam, mixtral, llama-moe} have emerged as a promising approach to mitigate this challenge. In MoE-based transformers, a single large FFN layer is replaced by an MoE layer, which consists of multiple smaller expert networks. During inference, instead of executing the entire set of parameters for an FFN layer, a router selects only a few relevant experts for each input token. This selective activation reduces the computation cost while achieving superior model performance. Applying MoE to edge devices \cite{edgemoe} directly translates to a reduction in EMA cost while reducing the overall computation. The weight EMA per token in the MoE model is substantially reduced compared to the conventional LLM, as shown in Equation~\eqref{eq:ema_moe}. $\mathbf{S}$ represents the number of deactivated experts and $\mathbf{P_{\text{expert}}}$ denotes the weight parameters of a single expert. As demonstrated by EdgeMoE \cite{edgemoe} and SMoLPU \cite{smolpu}, MoE models are well-suited for edge devices because they substantially reduce the inference cost of FFN layers.

%
\begin{equation}
\label{eq:ema_moe}
EMA_{\text{W, MoE}} = P_{\text{total}} - S \times P_{\text{expert}}
\end{equation}

Recent LLMs adopt speculative decoding (SD) \cite{medusa, 1st_sd, 2nd_sd, 3rd_sd, layerskip, bild, eagle3, quant-based} to reduce the repeated loading of model parameters. This is typically achieved by employing a smaller, low-cost draft model to generate multiple candidate tokens, which are then verified in a single pass by the larger target LLM model. The target model accepts only the verified tokens, which are denoted by accepted tokens. Such approaches accelerate decoding stages by producing multiple tokens for each stage. The application of speculative decoding is particularly beneficial for edge LLM devices \cite{edgellm, specmemo, smolpu}. Speculative decoding can significantly lower the EMA cost by reducing the frequent loading of large target model parameters. The weight EMA per token with speculative decoding is shown in Equation~\eqref{eq:ema_sd}. $\mathbf{N_{\text{A}}}$ represents the average number of accepted tokens per stage (acceptance length). EAGLE-3 \cite{eagle3} generates an average of five tokens per stage on the DeepSeek-R1 8B model \cite{deepseek-r1}, thereby reducing the model parameter load by approximately five times. Therefore, integrating speculative decoding with MoE models presents a compelling strategy for edge LLMs. This combination offers a twofold advantage: speculative decoding reduces the total number of decoding stages, while MoE reduces the EMA overhead of FFN layers within each stage.

%
\begin{equation}
\label{eq:ema_sd}
EMA_{\text{W, SD}} \approx P_{\text{total}} / N_{\text{A}}
\end{equation}
%

\begin{figure}
  \centering
  \begin{subfigure}[b]{\linewidth}
    \includegraphics[width=\linewidth]{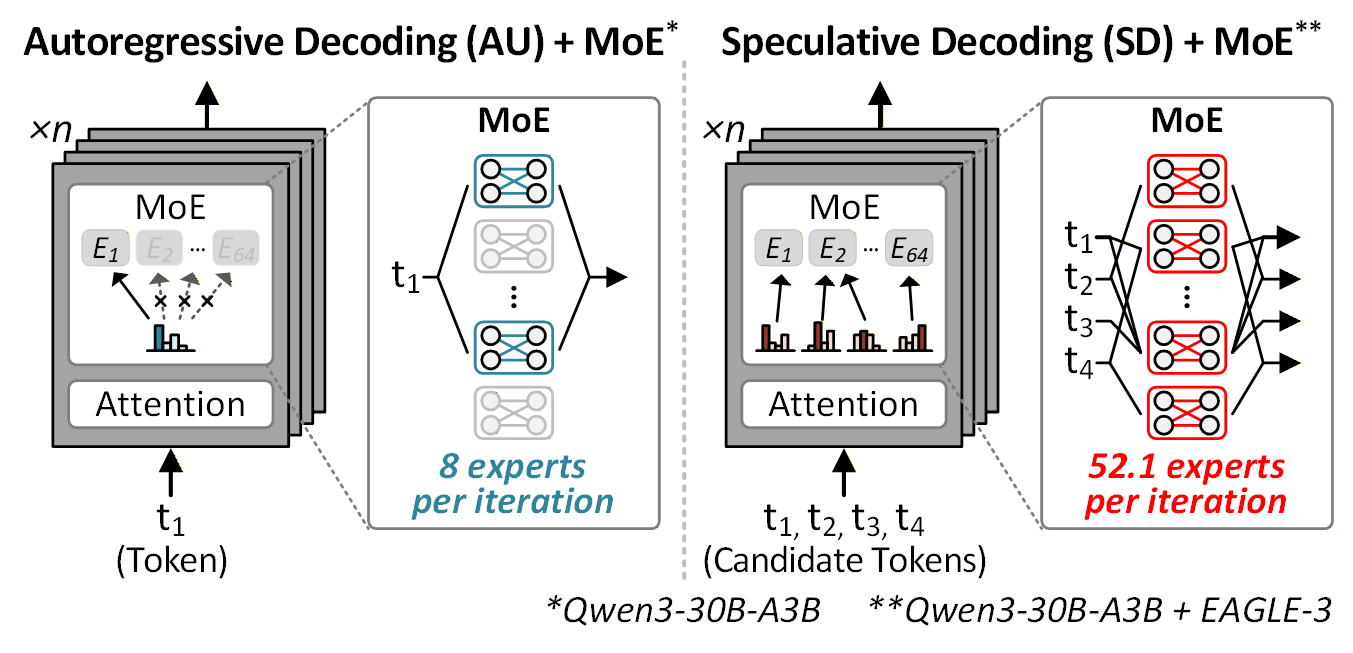}
    \caption{}\label{fig_2(a)}
  \end{subfigure}
  \hfill
  \begin{subfigure}[b]{\linewidth}
    \includegraphics[width=\linewidth]{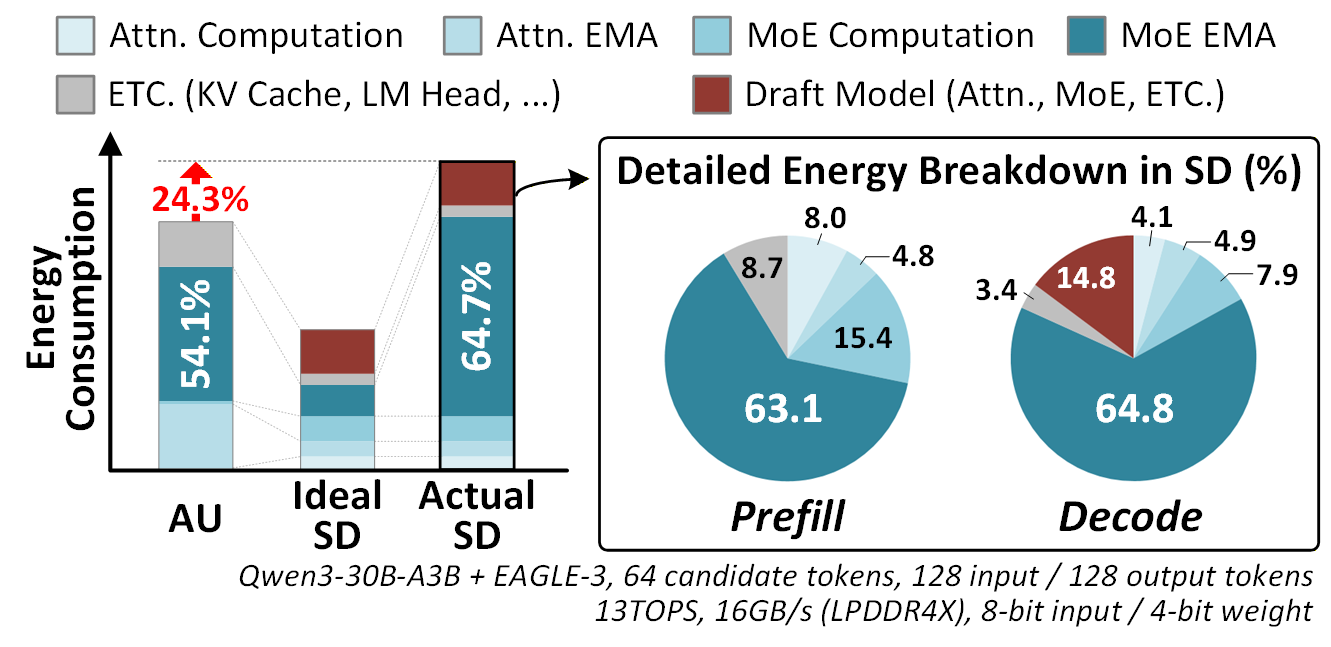}
    \caption{}\label{fig_2(b)}
  \end{subfigure}
  \hfill
  \captionsetup{justification=raggedright,singlelinecheck=false}
  \vspace{-2mm}
  \caption{(a) Incompatibility of speculative decoding and MoE.
  (b) Overall system breakdown.}
  \label{fig_2}
\end{figure}

However, naively applying both speculative decoding and MoE directly to edge devices incurs increased EMA overhead in MoE layers. Attaching EAGLE-3 \cite{eagle3} to Qwen3 \cite{qwen3} yields an average of 4.3 tokens per stage. Nonetheless, as shown in Figure~\ref{fig_2}(\subref{fig_2(a)}), each stage activates 52.1 experts, which increases EMA and computational cost by 6.5×. When combining two techniques, the weight EMA per token is approximated by Equation~\eqref{eq:ema_moesd}. Although increasing $\mathbf{N_{\text{A}}}$ and $\mathbf{S}$ is beneficial in minimizing EMA, these two factors present a significant trade-off. In speculative decoding, generating more candidate tokens increases the chance of accepted tokens ($\mathbf{N_{\text{A}}}$) and thus speeds up inference. In contrast, in an MoE model, a larger number of candidate tokens activates more experts per layer, which reduces $\mathbf{S}$ and lowers inference speed. Consequently, as noted in \cite{eagle1, moesd}, speculative decoding and MoE are incompatible.

%
\begin{equation}
\label{eq:ema_moesd}
EMA_{\text{W, MoE \& SD}} \approx (P_{\text{total}} - S \times P_{\text{expert}}) / N_{\text{A}}
\end{equation}

Figure~\ref{fig_2}(\subref{fig_2(b)}) presents the system energy breakdown when speculative decoding is applied to the MoE model. In the ideal case, speculative decoding (ideal SD) reduces the total energy by amortizing the energy of the target model over multiple accepted tokens. However, the actual result (actual SD) shows the opposite trend. While the EMA of the attention layer decreases, that of the MoE layer increases due to the incompatibility between speculative decoding and MoE. The draft model further introduces additional EMA and computation overhead, increasing the total energy consumption by 24.3\% compared to autoregressive decoding (AU). Since the draft model consists of only a single layer, the target model dominates the overall overhead. Speculative decoding increases computation by processing multiple candidate tokens simultaneously. However, the acceptance length ($\mathbf{N_{\text{A}}}$) saturates as the number of candidate tokens grows, meaning that arbitrarily increasing candidate tokens does not improve overall efficiency. Therefore, at the optimal number of candidate tokens, the EMA energy of the MoE layer remains the dominant bottleneck. The optimal number of candidate tokens will be discussed in Section~\ref{sec:SD}. As shown in the detailed energy breakdown, the MoE layer is the dominant bottleneck in both stages. In the prefill stage, many input tokens in the user prompt are processed together, activating a large number of experts. In the decode stage, the incompatibility between MoE and speculative decoding increases expert activation. As a result, the EMA of the MoE layer dominates the overall energy. This result does not imply that MoE and speculative decoding should be used separately. Instead, it shows that their memory incompatibility must be explicitly addressed. If this incompatibility is resolved, MoE and speculative decoding provide complementary benefits: MoE reduces the active FFN weights per decode stage, while speculative decoding reduces the number of decode stages.

To address the above EMA overhead of the MoE layer, we propose EdgeXpert, which combines software and hardware optimizations for memory-efficient LLM inference in edge devices. Experiments on speculative decoding \cite{eagle3} and MoE models \cite{qwen3, deepseek-v2, olmoe, granite} at various benchmarks show that EdgeXpert achieves up to 56.3\% latency reduction and 44.1\% energy reduction compared to prior works \cite{moe-pruner, edgemoe, smolpu}.

Key contributions of this work are as follows:
\begin{itemize}
\item{\textbf{\textit{Prefill Stage Optimization:}} We introduce prompt-wise expert reuse, which treats prefill routing not as independent per-token expert selection, but as prompt-level expert reuse using a shared expert set to reduce EMA while preserving routing quality. EdgeXpert first identifies important tokens using a lightweight encoder, builds the shared expert set from them, and then applies budgeted expert routing to less important tokens. Furthermore, the partitioning network supports variable-length input tokens with negligible latency overhead. Together, these techniques substantially reduce the EMA of MoE layers in the prefill stage.}
\item{\textbf{\textit{Decode Stage Optimization:}} We propose a depth-aware expert coalescing that reduces EMA during the decode stage. Unlike existing expert pruning, our approach leverages the depth-wise mutual exclusivity inherent to candidate tokens to load only the necessary expert channels. This achieves EMA reductions that are difficult to achieve through existing pruning strategies alone.}
\end{itemize}
\section{Background}
\label{sec:background}

\begin{figure}[t]
  \centering
    \hspace*{\fill} 
    \includegraphics[width=0.95\linewidth]{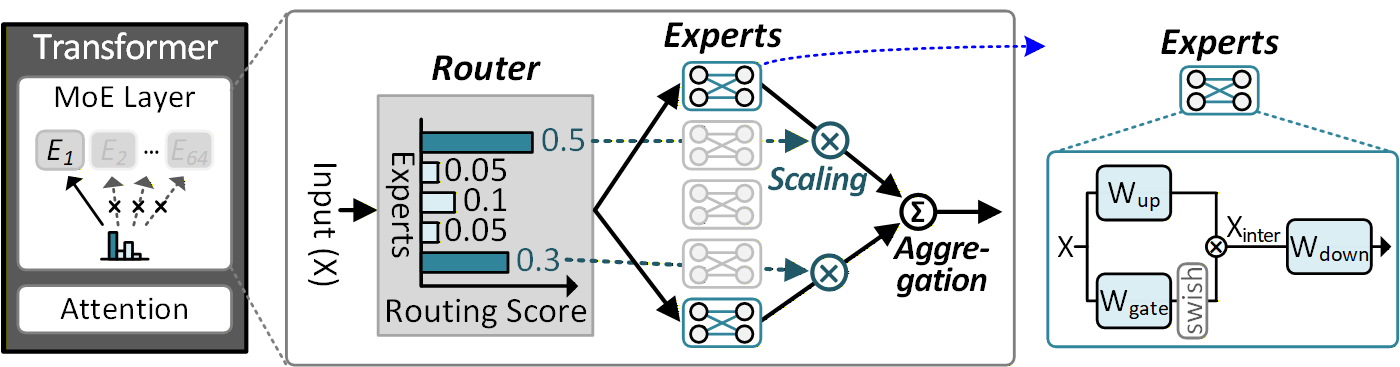}
    \hspace*{\fill} 
  \caption{Operational flow of MoE.}
  \vspace{-1mm}
  \label{fig_3}
\end{figure}

\subsection{Mixture-of-Experts}
The MoE model achieves high capacity with minimal computation by routing each input token to a small subset of specialized experts rather than executing all experts. In a typical MoE layer (Figure~\ref{fig_3}), a router computes routing scores, indicating the selection probability of each expert. The input selects top-k experts and then performs the GEMV operation with the selected experts. Each expert’s output is scaled by its routing score and aggregated into the final result.

Prior MoE works have focused on reducing expert loading latency. \cite{cache-prior, moe_alg2} optimize the routing policy of the router to increase the expert hit rate and thereby lower loading latency. However, these works are infeasible on edge devices due to stringent on-chip memory capacity constraints \cite{c-transformer, mecla, broca}, where the available cache size is typically on the order of a few megabytes (MB). For instance, a single expert in DeepSeek-V2-Lite \cite{deepseek-v2} occupies 4 MB at 4-bit precision, making multi-expert caching and scheduling impractical. Therefore, each expert must be loaded sequentially, and all routed tokens must be computed in parallel before loading the next expert.

Several MoE hardware studies~\cite{duplex, space-mate} improve the efficiency of expert execution under unbalanced routing. Duplex~\cite{duplex} uses a heterogeneous xPU-PIM architecture to accelerate memory-bound MoE operations, reducing the cost of memory access through PIM offloading. Space-Mate~\cite{space-mate} improves PE utilization by simultaneously executing high-reusability and low-reusability experts. These schemes improve how activated experts are served, but they do not reduce the number of experts activated per request. 

\begin{figure}
  \centering
  \begin{subfigure}[b]{\linewidth}
    \includegraphics[width=\linewidth]{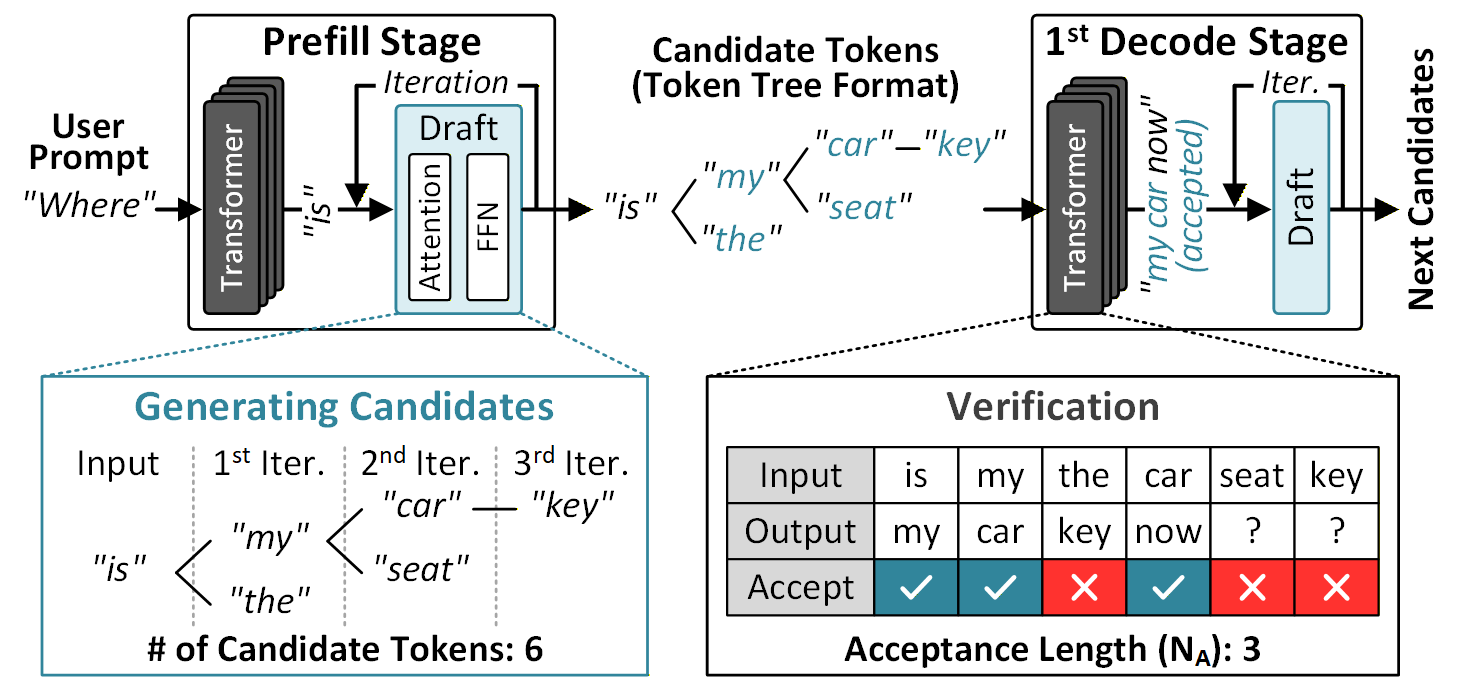}
    \caption{}\label{fig_4(a)}
  \end{subfigure}
  \hfill
  \begin{subfigure}[b]{\linewidth}
    \includegraphics[width=\linewidth]{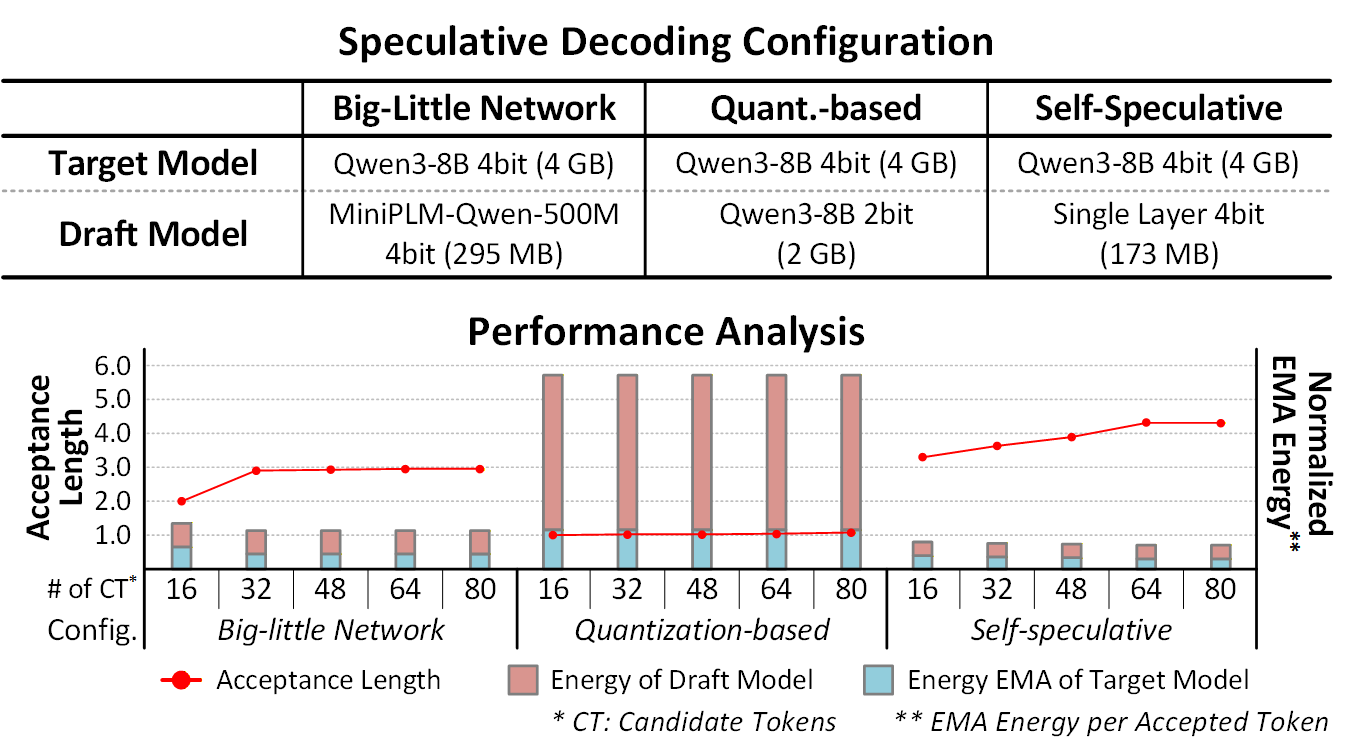}
    \caption{}\label{fig_4(b)}
  \end{subfigure}
  \hfill
  \captionsetup{justification=raggedright,singlelinecheck=false}
  \vspace{-4mm}
  \caption{(a) Operational flow of speculative decoding.
  (b) Comparison of various speculative decoding frameworks.}
  \label{fig_4}
\end{figure}

\subsection{Speculative Decoding}
\label{sec:SD}
Speculative decoding was proposed to overcome the limitations of autoregressive decoding. By using a small draft model to speculate future tokens and verifying them in parallel with the target model, overall decoding latency is reduced while preserving output quality.

Recent speculative decoding approaches fall into three categories. First, the big-little model \cite{bild} is the most fundamental approach. A small draft model generates multiple candidate tokens with little overhead. Candidate tokens are then fed in parallel into the larger target model for verification. C-Transformer \cite{c-transformer} applies this scheme on edge devices, improving energy efficiency. However, to maintain a high acceptance length (speculation accuracy), the draft model must be large enough (i.e., above a certain parameter count), making it impractical for edge devices.

Second, quantization-based speculative decoding \cite{quant-based} uses a low-bit quantized model as the draft model, while a higher-precision model is used for verification. However, on resource-constrained edge devices, the target model is already deployed at low precision \cite{broca, c-transformer, smolpu}. In this setting, it is difficult to further reduce the precision of the draft model because the draft model has to use even lower precision than the already quantized target model. For example, if the target model is quantized to 4-bit, even an aggressively quantized 2-bit draft model still has half the footprint of the target model. As a result, the draft model still incurs a large EMA, which limits the benefit on edge devices.

Third, self-speculative decoding \cite{medusa,eagle1, eagle3} attaches a lightweight transformer block to the tail of the target model itself as a draft model. As illustrated in Figure~\ref{fig_4}(\subref{fig_4(a)}), input tokens first pass through the target model. Subsequently, the draft model runs iteratively to generate multiple candidate tokens, which are then structured into a token tree. This tree serves as the input for the next verification pass through the target model. Because only one transformer block is used iteratively for speculation, the cost remains low even on edge devices. Moreover, the speculative block is fine-tuned to align its parameters with the target model, yielding a high acceptance length. Thus, self-speculative decoding is preferable for edge devices.

Figure~\ref{fig_4}(\subref{fig_4(b)}) compares the three approaches in terms of acceptance length and the EMA energy under the same verification model. Normalized EMA energy is reported per accepted token. It includes the repeated draft-model EMA required to generate all candidate tokens and the target-model EMA for one verification pass, normalized by the resulting acceptance length. For a fair comparison, the same draft model size is used across approaches. The draft model of the quantization-based approach cannot be further compressed and therefore requires a larger draft model than the other two. The big-little approach achieves a low acceptance length due to the limited capacity of its small draft model. The quantization-based approach has a very short acceptance length due to 2-bit quantization. Although using a 3-bit draft model can increase the acceptance length, the draft model becomes too large. The EMA of the 2-bit draft model alone exceeds the combined EMA of the target and draft models in the other approaches, resulting in the highest total EMA ratio. Self-speculative decoding achieves a competitive acceptance length with a low EMA, making it the most suitable approach for edge devices. Based on this analysis, the optimal number of candidate tokens is set to 64.


Specialized hardware for speculative decoding has emerged. SpecInfer \cite{specinfer} introduces tree attention for parallel verification. EdgeLLM~\cite{edgellm} dynamically builds the token tree and continues drafting during verification. However, these works focused on improving acceptance length or reducing verification latency, while overlooking EMA overhead in edge scenarios.



\makeatletter
\def\ymark{\textcolor{green!55!black}{$\checkmark$}}
\def\nmark{\textcolor{red}{$\times$}}
\def\pmark{\textcolor{orange}{$\triangle$}}
\def\lowmark{\textcolor{green!55!black}{+}}
\def\medmark{\textcolor{orange}{++}}
\def\highmark{\textcolor{red}{+++}}
\makeatother

\begin{table}[!t]
\centering
\caption{Comparison with speculative decoding works\\ for MoE models}
\label{table_0}
\footnotesize                          
\setlength{\tabcolsep}{2pt}            
\renewcommand{\arraystretch}{1.15}
\begin{tabular}{@{}c|c|c|c|c|c@{}}
\toprule
 & \textbf{Optimized} & \textbf{Drafting} & \textbf{Prefill EMA}
 & \multicolumn{2}{c}{\textbf{Decode EMA}} \\
\cmidrule(lr){4-6}
 & \textbf{Model} & \textbf{Overhead} & \textbf{\textit{Expert}}
 & \textbf{\textit{Expert}} & \textbf{\textit{Intra-Expert}} \\
\midrule
\textbf{MoESD}     & Target        & \medmark  & \nmark & \pmark & \nmark \\
\textbf{SS-MoE}    & Draft, Target & \highmark & \nmark & \pmark & \nmark \\
\textbf{MoE-Spec}  & Target        & \lowmark  & \nmark & \ymark & \nmark \\
\textbf{SMoLPU}    & Target        & \lowmark  & \nmark & \ymark & \pmark \\
\textbf{EdgeXpert} & Target        & \lowmark  & \ymark & \ymark & \ymark \\
\bottomrule
\end{tabular}

\vspace{2pt}
{\footnotesize \raggedleft
\lowmark: low,\quad \medmark: medium,\quad \highmark: high\par}
\end{table}

\subsection{Related Works}
Applying speculative decoding to MoE models yields suboptimal performance gains. To address this, existing pruning schemes can be leveraged to reduce redundant parameters in this system.

\textbf{\textit{Candidate Token Pruning.}} During speculative decoding, candidate tokens that are incorrectly speculated are discarded after verification by the target model. The computation spent on these discarded tokens is entirely redundant. Candidate token pruning \cite{c_prun1, c_prun2} removes low-confidence candidates before verification, but its benefit decreases as speculative decoding becomes more accurate. In our Qwen3~\cite{qwen3} evaluation, early frameworks such as Medusa~\cite{medusa} allow pruning 74.6\% candidate tokens at 20\% $\mathbf{N_A}$ degradation. However, recent frameworks~\cite{eagle3, sequoia} achieve substantially higher speculation accuracy. Only 13.5\% of candidate tokens in EAGLE-3 can be pruned at the same degradation point. Therefore, candidate token pruning yields diminishing returns and is no longer an effective strategy for reducing inference overhead.

\textbf{\textit{Expert Pruning.}} Recent works \cite{e_prun1, e_prun2, e_prun3, moe-pruner} have shown that not all activated experts contribute meaningfully to model output. They prune less important experts or specific channels within experts to reduce overhead. More recently, SMoLPU \cite{smolpu} presents a state-of-the-art pruning framework for MoE-based speculative decoding and serves as a natural baseline for our study. First, it predicts and prunes experts that are activated exclusively by tokens expected to be discarded after verification. Second, it performs fine-grained channel pruning within each activated expert using a dynamic policy based on the channel importance of input tokens. This coarse-to-fine pruning removes a substantial number of parameters with minimal accuracy loss. Because SMoLPU already demonstrates effective pruning in a system that combines MoE and speculative decoding, we adopt it as the baseline pruning scheme. However, SMoLPU has two fundamental limitations that leave the primary bottleneck unresolved. First, its pruning mainly reduces computation rather than EMA. Second, it focuses only on the decode stage and does not consider the prefill stage, where multiple input tokens (user prompt) can simultaneously activate many experts. EdgeXpert adopts the coarse-to-fine pruning of SMoLPU as a baseline and introduces EMA-aware optimizations across both stages to address these unresolved bottlenecks.

\begin{figure}[t]
  \centering
    \hspace*{\fill} 
    \includegraphics[width=0.95\linewidth]{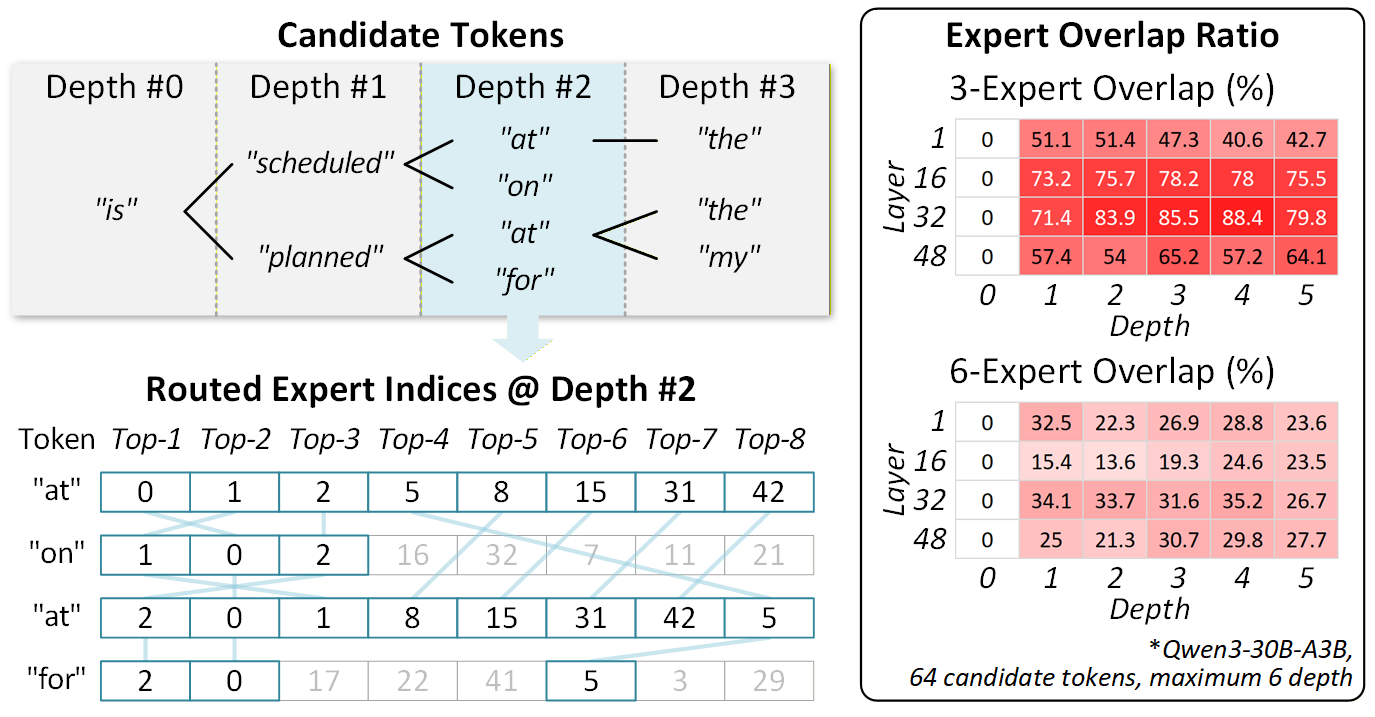}
    \hspace*{\fill} 
    \caption{Contextual similarity and expert overlap ratio.}
  \label{fig_5}
\end{figure}

\begin{figure*}[t]
  \centering
    \hspace*{\fill} 
    \includegraphics[width=\linewidth]{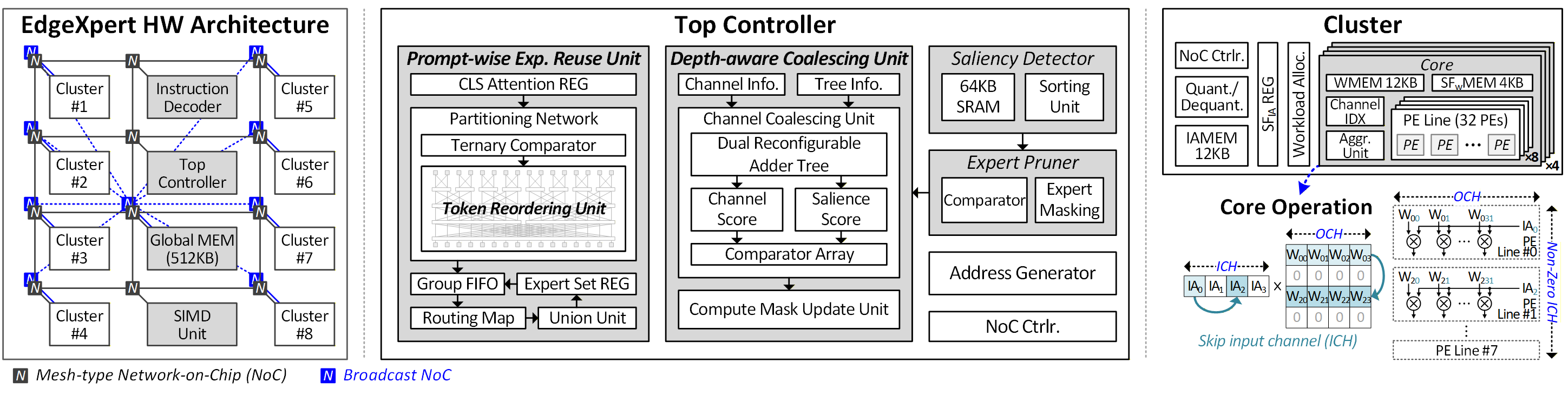}
    \hspace*{\fill} 
  \vspace{-4mm}
  \caption{Overall architecture.}
  \label{fig_6}
\end{figure*}

\textit{\textbf{MoE with speculative decoding.}} Recent works~\cite{moesd, ss-moe, moe-spec, smolpu} have studied speculative decoding for MoE models. Table~\ref{table_0} summarizes their optimization coverage. MoESD~\cite{moesd} analyzes the interaction between MoE and speculative decoding on GPUs and searches for an efficient operating point by adjusting parameters such as batch size and candidate-token configuration. It does not directly reduce the expert EMA introduced by candidate tokens. SS-MoE~\cite{ss-moe} constructs the draft path from the same target MoE model by activating only the top-1 routed expert, instead of using a separate draft model. However, this self-draft path remains costly for edge deployment. For DeepSeek-V2-Lite, top\nobreakdash-1 expert routing still activates about 1.3B parameters, while the EAGLE-3 draft block used in EdgeXpert contains only 351M parameters~\cite{eagle3}. Moreover, SS-MoE mainly reduces CPU-GPU expert transfers, rather than reducing the expert EMA. MoE-Spec~\cite{moe-spec} adopts EAGLE-3 and reduces verification overhead through expert budgeting. This effectively limits expert-level EMA during decode, but its budgeted routing still does not address intra-expert EMA. SMoLPU~\cite{smolpu} applies coarse-to-fine pruning during the decode stage. However, its fine-grained pruning mainly reduces computation and does not fully translate into EMA reduction. In contrast, EdgeXpert addresses both prefill expert EMA and decode intra-expert EMA, providing broader coverage for edge deployment.
\section{Motivation}
\label{sec:motivation}

\subsection{CLS Token as a Global Context Summarizer}
The CLS token is a special token attached to the input sequence in encoder-based transformers such as BERT \cite{bert}. It is trained to summarize the entire input into a single representation. In vision transformers \cite{vit}, the CLS token captures semantically important regions, and in retrieval-augmented generation (RAG) \cite{rag}, the CLS token embedding represents the entire document for similarity search. This property makes the CLS token an effective lightweight estimator of token importance. Recent VLM works such as FasterVLM~\cite{fastervlm}, HiPrune~\cite{hiprune}, and ATP-LLaVA~\cite{atp-llava} also exploit CLS tokens or attention scores to identify important visual tokens for pruning. These works use token importance to remove less important visual tokens and shorten the input sequence. Unlike these works, EdgeXpert uses CLS tokens to estimate the importance of input tokens and determine which tokens should create or reuse experts to reduce expert EMA.

\subsection{Characteristics of Candidate Tokens}
\textit{\textbf{Depth-wise Contextual Similarity.}}
Figure~\ref{fig_5} illustrates the contextual similarity among candidate tokens generated at the same depth. Candidate tokens are commonly represented as a token tree. Tokens at the same depth exhibit a high probability of sharing grammatical or contextual characteristics. This similarity is validated by examining the overlap ratio of activated experts. Our analysis shows that for tokens at the same depth, the co-activation ratio for three or more experts among eight experts is up to 88.4\%. This observation is consistent with Figure~\ref{fig_2}(\subref{fig_2(b)}). Speculative decoding increases the total number of activated experts by verifying multiple candidate tokens, but many of these activations overlap within the same depth group. Thus, MoE-based speculative decoding increases the overall expert activations while also creating depth-wise expert reuse opportunities. EdgeXpert exploits this overlap to reduce redundant expert EMA during verification.

\textit{\textbf{Depth-wise Mutual Exclusivity.}}
In addition to contextual similarity, candidate tokens at the same depth also demonstrate mutual exclusivity. Within a single depth, only one token is accepted during verification, and other tokens are discarded. This characteristic presents an opportunity for hardware-efficient computation. It enables precise computations for the token likely to be accepted, while using approximate computations for the remaining candidates.
\section{EdgeXpert: An Edge Device for LLM Inference}
\label{sec:EdgeXpert}

Prior work \cite{smolpu} on MoE-based speculative decoding has demonstrated that redundant experts can be effectively pruned through a coarse-to-fine strategy during the decode stage. We adopt this pruning scheme as our baseline. However, the dominant bottleneck on edge devices is EMA rather than computation. Pruning introduces sparsity and reduces computation, but it does not proportionally reduce EMA. We propose EdgeXpert to address this limitation by minimizing EMA across both prefill and decode stages.

However, without hardware-level support, sparsity does not directly translate into EMA reduction. Since EdgeXpert dynamically determines the required experts and channels from the input tokens at runtime, the memory system must support on-the-fly fine-grained expert/channel loading before unnecessary data is fetched. This requirement is especially important in the decode stage, where the channel set to be loaded is dynamically constructed from same-depth candidate tokens. Therefore, EdgeXpert adopts a SW-HW co-design that skips unnecessary expert/channel data before EMA occurs. EdgeXpert achieves higher performance than prior software-centric works~\cite{moe-pruner, edgemoe}, which determine pruning masks or bit-precision offline using a validation set before inference.


\subsection{Overall Architecture}
Figure~\ref{fig_6} shows the overall architecture. EdgeXpert comprises an instruction decoder, a top controller, a 512 KB global SRAM, a SIMD unit, and eight clusters. The PE array contains 8K MACs, comparable to recent edge LLM accelerators~\cite{smolpu, broca, c-transformer}, while the 512 KB SRAM is sufficient to buffer input activations, routing metadata, masks, and streamed weight blocks for the proposed dataflow. Since edge accelerators typically provide only a few MB of on-chip SRAM and each MoE expert requires several MB, caching and reusing multiple experts on-chip is impractical. Thus, EdgeXpert uses 512 KB SRAM because increasing the SRAM capacity would add area and power while still being insufficient to keep multiple experts resident on chip. Each expert is fetched once from external memory, kept on-chip until all routed tokens are processed, and then replaced by the next expert. The datapath adapts to the distinct characteristics of the prefill and decode stages. In the prefill stage, each expert is reused across many tokens, resulting in high operational intensity (Op/B). EdgeXpert exploits this high reusability through a broadcast NoC that distributes the same weight to all clusters simultaneously. In contrast, in the decode stage, different output channels are assigned to each cluster. Experts are loaded from external memory along the output-channel dimension. Since channel pruning is applied along the input-channel dimension, this loading order preserves contiguous memory accesses and avoids fragmented DRAM transactions.

\begin{figure}
  \centering
  \begin{subfigure}[b]{\linewidth}
    \includegraphics[width=\linewidth]{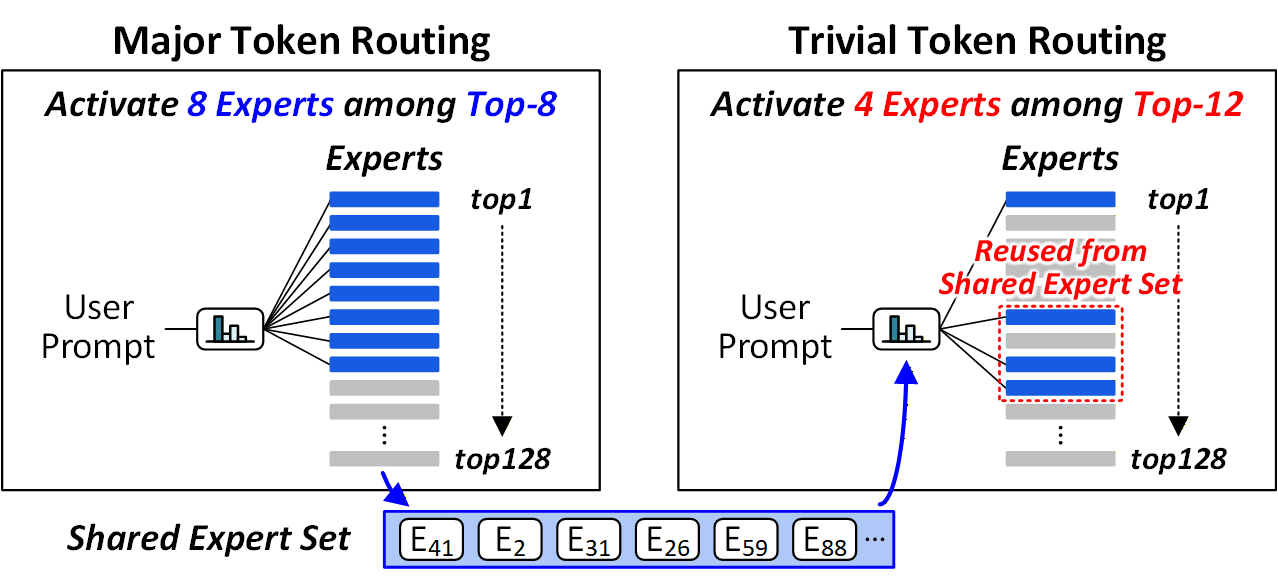}
    \caption{}\label{fig_7(a)}
  \end{subfigure}
  \hfill
  \begin{subfigure}[b]{\linewidth}
    \includegraphics[width=\linewidth]{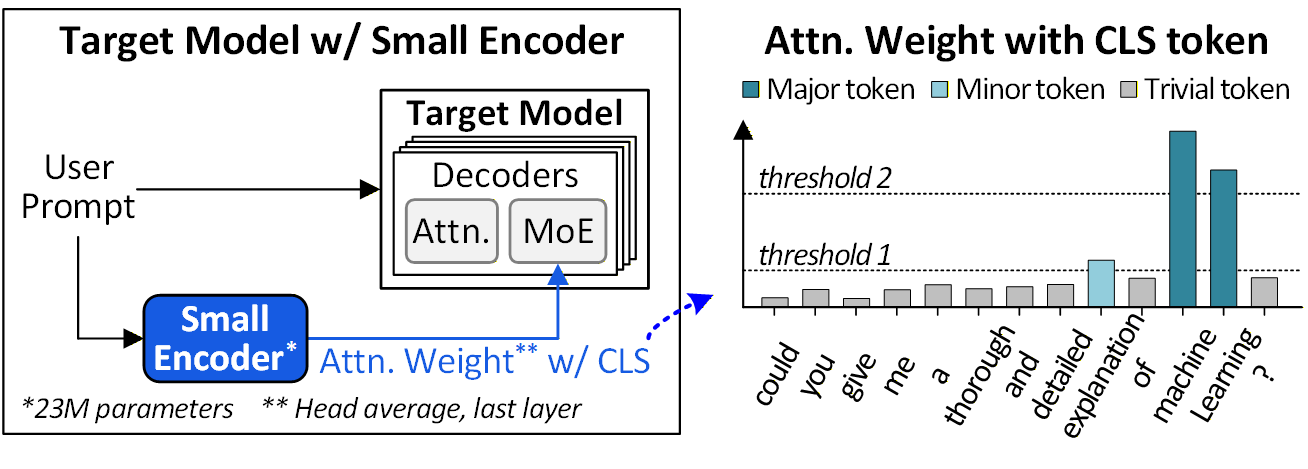}
    \caption{}\label{fig_7(b)}
  \end{subfigure}
  \hfill
  \begin{subfigure}[b]{\linewidth}
    \centering
    \includegraphics[width=\linewidth]{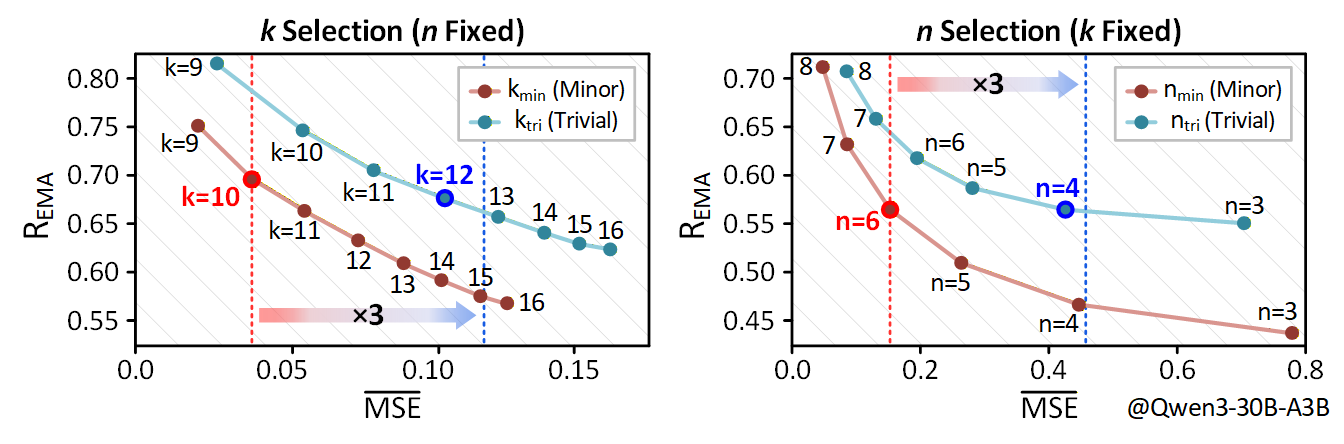}%
    \caption{}\label{fig_7(c)}
  \end{subfigure}
  \hfill
  \begin{subfigure}[b]{\linewidth}
    \includegraphics[width=\linewidth]{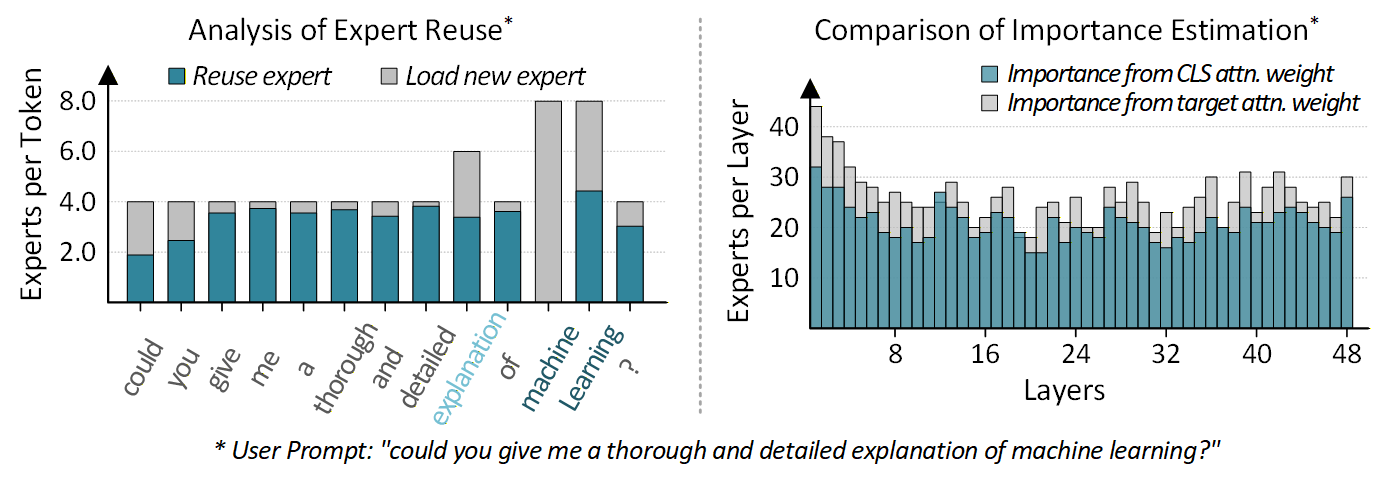}
    \caption{}\label{fig_7(d)}
  \end{subfigure}
  \hfill
  \captionsetup{justification=raggedright,singlelinecheck=false}
  \vspace{-3mm}
  \caption{(a) Example of shared expert reuse.
  (b) Operation flow in the prefill stage.
  (c) Example to determine optimal $k$ and $n$.
  (d) Analysis of expert reuse and comparison of importance estimation.}
  \vspace{-4mm}
  \label{fig_7}
\end{figure}

The top controller orchestrates prompt-wise expert reuse and expert channel coalescing. The prompt-wise expert reuse unit manages the routing of the prefill stage. It contains a partitioning network, a group FIFO, a union unit, and a shared expert set register. The partitioning network uses ternary comparators and a token reordering unit to classify input tokens into importance groups based on CLS attention. The group FIFO and the union unit then form and update the shared expert set and generate routing decisions for each token group. The depth-aware coalescing unit operates during the decode stage to reduce EMA when loading pruned experts. The channel coalescing unit evaluates per-channel importance and masks channels. The compute mask update unit adjusts channel masks to recover the accuracy drop caused by coalescing.

The cluster is composed of a quantization/dequantization unit, a workload allocator, a 12KB input memory ($\mathrm{IAMEM}$), and four cores. The workload allocator improves utilization under imbalanced data reuse (Op/B) by allocating input channels to each core. Each core consists of an input buffer, a 12KB weight memory ($\mathrm{WMEM}$), a 4KB weight scaling factor memory ($\mathrm{SF}_\mathrm{W}\mathrm{MEM}$), eight PE lines, an aggregation unit, and the nonzero channel index register, which enables efficient input channel skipping. Furthermore, each PE line comprises 32 PEs and supports sparse computation via channel-skipping operations. Each PE line is mapped along the output channel dimension (OCH), where weights are unicast to individual PEs while a single input is broadcast across all 32 PEs within the line. Across PE lines, the mapping follows the input channel (ICH) dimension, where only non-zero inputs are fed to the respective PE lines, effectively skipping computation for pruned channels. The 32 partial products from each PE line are forwarded to adder trees. Adder tree outputs are then accumulated in the aggregation unit.


\begin{algorithm}[t]
\footnotesize\raggedright
\caption{Prompt-wise Expert Reuse Policy}
\label{alg:routing-policy}
\begin{algorithmic}[1]
    \Statex \hspace{-\algorithmicindent} \textbf{Input:}
    \Statex $\mathcal{T}^{maj}, \mathcal{T}^{min}, \mathcal{T}^{tri}$: Major, minor, and trivial token sets.
    \Statex $I^{maj}, I^{min}, I^{tri}$: Top-$k_{maj}$, Top-$k_{min}$, Top-$k_{tri}$ expert indices.
    \Statex $n_{maj}, n_{min}, n_{tri}$: number of routed experts of $\mathcal{T}^{maj}, \mathcal{T}^{min}, \mathcal{T}^{tri}$.
    \Statex \hspace{-\algorithmicindent} \textbf{Output:}
    \Statex $\mathcal{E}$: Shared expert set.
    \Statex $\mathcal{R}^{maj}, \mathcal{R}^{min}, \mathcal{R}^{tri}$: Final expert indices for tokens.
    \Statex
    \State // 1. Major token routing
    \For{each $t \in \mathcal{T}^{\mathrm{maj}}$}
        \State $\mathcal{R}_t^{maj} \leftarrow I_t^{maj}[1{:}n_{maj}]$
        \State $\mathcal{E} \leftarrow \mathcal{E} \cup \mathcal{R}_t^{maj}$
    \EndFor
    \State
    \State // 2. Minor token routing
    \For{each $t \in \mathcal{T}^{\mathrm{min}}$}
        \State $\mathcal{R}_t^{min} \leftarrow I_t^{min}[1]$
        \State $\mathcal{R}_t^{min} \leftarrow \mathcal{R}_t^{min} \cup reuse\_route(I_t^{min}[2{:}k_{min}], \mathcal{E}, n_{min}-1)$
        \If{$|\mathcal{R}^{min}_t| < n_{min}$}
                \State $\mathcal{R}_t^{min} \leftarrow \mathcal{R}_t^{min} \cup top\_k\_route(I_t^{min}, \mathcal{R}_t^{min}, n_{min} - |\mathcal{R}^{min}_t|)$
            \EndIf
        \State $\mathcal{E} \leftarrow \mathcal{E} \cup \mathcal{R}_t^{min}$
    \EndFor
    \State
    \State // 3. Trivial token routing
    \For{each $t \in \mathcal{T}^{\mathrm{tri}}$}
        \State $\mathcal{R}_t^{tri} \leftarrow I_t^{tri}[1]$
        \State $\mathcal{R}_t^{tri} \leftarrow \mathcal{R}_t^{tri} \cup reuse\_route(I_t^{tri}[2{:}k_{tri}], \mathcal{E}, n_{tri}-1)$
        \If{$|\mathcal{R}^{tri}_t| < n_{tri}$}
                \State $\mathcal{R}_t^{tri} \leftarrow \mathcal{R}_t^{tri} \cup top\_k\_route(I_t^{tri}, \mathcal{R}_t^{tri}, n_{tri} - |\mathcal{R}^{tri}_t|)$
            \EndIf
    \EndFor
    \State
    \State \Return $\mathcal{E}, \mathcal{R}^{maj}, \mathcal{R}^{min}, \mathcal{R}^{tri}$
    \State 
    \Function{$reuse\_route$}{$I, \mathcal{E}, m$} \ \ \ \ \ \textbf{\textit{// Reuse shared expert set}}
    \State \Return top-$m$ experts in $I \cap \mathcal{E}$
    \EndFunction 
    \State
    \Function{$top\_k\_route$}{$I, \mathcal{R}, m$} \ \ \ \ \ \textbf{\textit{// Fill remaining experts}}
    \State \Return top-$m$ experts from \(I \setminus \mathcal{R}\)
    \EndFunction

\end{algorithmic}
\end{algorithm}

\subsection{Prompt-wise Expert Reuse}

In the prefill stage, routing every token independently activates too many experts and increases expert EMA. To address this inefficiency, we propose prompt-wise expert reuse, a prefill stage routing policy that reduces EMA by allowing less important tokens to reuse experts already activated by more important tokens. Figure~\ref{fig_7}(\subref{fig_7(a)}) illustrates the key intuition. Important tokens (major) establish a shared expert set, while less important tokens (minor/trivial) preferentially reuse that set instead of activating new experts.

EdgeXpert first uses a lightweight encoder~\cite{minilm} to identify important tokens. The encoder introduces only negligible overhead, while substantially reducing expert EMA. As shown in Figure~\ref{fig_7}(\subref{fig_7(b)}), EdgeXpert uses the CLS attention weights produced by the lightweight encoder to partition tokens into major, minor, and trivial groups and then applies different routing budgets to each group.

Algorithm~\ref{alg:routing-policy} describes the budgeted expert routing policy for prompt-level expert reuse. For each token group, EdgeXpert considers the top-$k$ candidate experts and selects $n$ routed experts. Major tokens use the original routing budget, i.e., $k_{maj}$ and $n_{maj}$ follow the original router configuration, and their routed experts form shared expert set $\mathcal{E}$. Minor and trivial tokens use a larger candidate range but a smaller routing budget, i.e., $k_{tri} > k_{min} > k_{maj}$ = $n_{maj} > n_{min} > n_{tri}$, to increase reuse while reducing newly activated experts. For minor tokens, EdgeXpert preserves the top-1 expert to keep the dominant routing path. The remaining $n_{min}-1$ slots are first filled by \texttt{reuse\_route} using experts in $\mathcal{E}$. If they are insufficient, \texttt{top\_k\_route} fills the remaining slots with unselected experts in descending score order. The selected minor-token experts are then merged into $\mathcal{E}$. Trivial tokens follow the same rule with the updated $\mathcal{E}$ and the more aggressive budget $(k_{tri}, n_{tri})$.

We determine \((k,n)\) by searching for a Pareto point between the MoE-layer error ($\overline{\mathrm{MSE}}$) and the activated expert count ($R_{\mathrm{EMA}}$). For each \((k,n)\), we compare the final MoE-layer output of the proposed routing policy with that of the original routing, and measure the activated expert count (Equation~\eqref{k_n_policy}).
\begin{equation}
\label{k_n_policy}
\begin{aligned}
\overline{\mathrm{MSE}}(k,n) &= \mathbb{E}_{\ell,t}\!\left[
\frac{\|h_{\ell,t}^{\mathrm{ori}}-h_{\ell,t}^{k,n}\|_2^2}
{\|h_{\ell,t}^{\mathrm{ori}}\|_2^2}\right], \\[2pt]
R_{\mathrm{EMA}}(k,n) &= \frac{\sum_{\ell}|E_{\ell}^{k,n}|}
{\sum_{\ell}|E_{\ell}^{\mathrm{ori}}|}
\end{aligned}
\end{equation}
\(\ell\) and \(t\) denote the layer and token indices, respectively. \(h_{\ell,t}^{\mathrm{ori}}\) and \(h_{\ell,t}^{k,n}\) are the MoE-layer outputs of the original routing and the proposed \((k,n)\) routing, respectively. \(E_{\ell}^{k,n}\) is the activated expert-set size per layer. During the search, we iteratively adjust \(k\) and \(n\). \(k\) is increased to expand reusable expert candidates, and \(n\) is decreased to reduce activated experts. At each step, we select the elbow point between \(\overline{\mathrm{MSE}}\) and \(R_{\mathrm{EMA}}\). As shown in Figure~\ref{fig_7}(\subref{fig_7(c)}), for trivial tokens, we allow more aggressive reuse and choose the lowest $R_{\mathrm{EMA}}$ point whose $\overline{\mathrm{MSE}}$ is within \(3\times\) the minor-token elbow-point $\overline{\mathrm{MSE}}$.

Prior MoE routing methods \cite{cache-prior, pre-gated_moe} treat routing as a token-wise expert selection problem and mainly optimize locality and latency. In contrast, EdgeXpert treats routing as a prompt-level expert set formation problem. EdgeXpert directly reduces the EMA of the prefill stage and improves both energy consumption and latency.

As shown in Figure~\ref{fig_7}(\subref{fig_7(d)}), important tokens add new experts to the shared expert set, while less important tokens predominantly reuse experts from this shared set. We further compared our method with an alternative scheme that estimates token importance from the attention weights of the target model. Although this scheme uses the target model directly, the causal attention mask inherently restricts each token to attending only to preceding tokens, limiting its ability to accurately assess importance across the full context. By contrast, our lightweight encoder leverages bidirectional attention to compute token importance before the target model is executed, suppressing expert EMA from the beginning of the prefill stage. 

\begin{figure}
  \centering
  \begin{subfigure}[b]{\linewidth}
    \includegraphics[width=\linewidth]{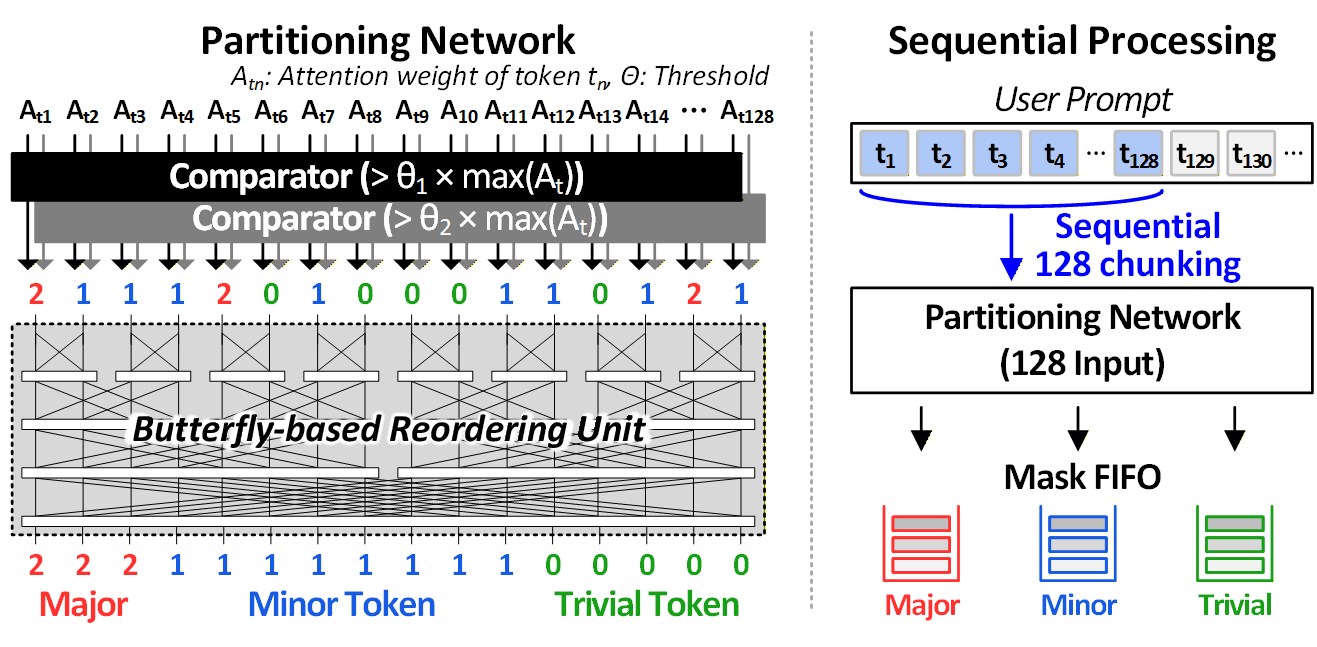}
    \caption{}\label{fig_8(a)}
  \end{subfigure}
  \hfill
  \begin{subfigure}[b]{\linewidth}
    \includegraphics[width=\linewidth]{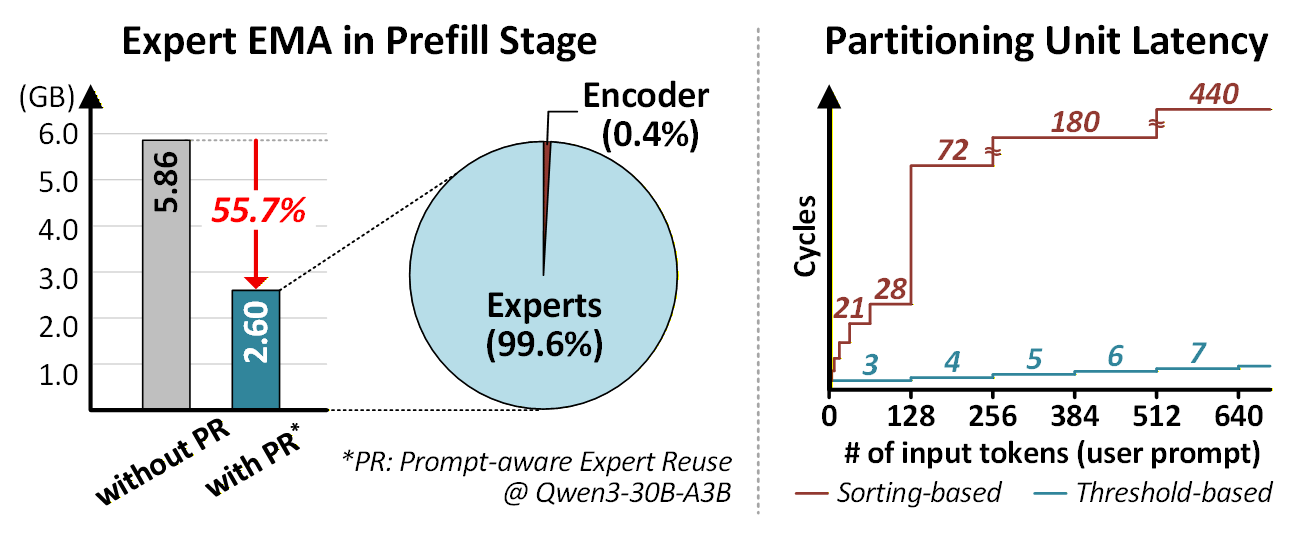}
    \caption{}\label{fig_8(b)}
  \end{subfigure}
  \hfill
  \captionsetup{justification=raggedright,singlelinecheck=false}
  \vspace{-4mm}
  \caption{(a) Detailed hardware of the partitioning network.
  (b) Performance analysis of prompt-wise expert reuse.}
  \vspace{-3mm}
  \label{fig_8}
\end{figure}

Figure~\ref{fig_8}(\subref{fig_8(a)}) shows the hardware implementation of the proposed policy. The partitioning network compares the CLS attention weight of each token against two thresholds using parallel comparator arrays, generating a ternary label for every token. These labels are then reordered by a butterfly-based reordering unit to partition the input into major, minor, and trivial groups. The proposed hardware processes 128 input tokens per cycle and is implemented as a three-stage pipeline, enabling low-latency token grouping.

A straightforward alternative is to sort tokens globally by attention weight and partition them by rank. However, such a sorting-based design is inefficient for the variable-length user prompt. Once the prompt length exceeds the hardware input width, a sorting-based design requires chunk-wise sorting followed by cross-chunk merging, causing latency to grow rapidly with input length. In contrast, the proposed threshold-based partitioning network avoids global ordering and processes each 128-token block independently, making the design suitable for variable-length prefill inputs.

Figure~\ref{fig_8}(\subref{fig_8(b)}) shows the effectiveness of the proposed method and its hardware support. By allowing less important tokens to reuse the shared expert set, EdgeXpert reduces expert EMA by 55.7\% in the prefill stage. Importantly, the additional EMA introduced by the lightweight encoder used for importance estimation is negligible, indicating that the reduction in expert EMA substantially outweighs the encoder overhead. Moreover, as the input length increases, the proposed threshold-based design exhibits near-linear latency scaling, whereas a sorting-based implementation incurs rapidly increasing latency due to repeated sorting and cross-chunk merging. These results show that prompt-wise expert reuse is effective.


\subsection{Depth-aware Expert Coalescing}

\begin{figure}[t]
  \centering
  \begin{subfigure}[b]{\linewidth}
    \includegraphics[width=\linewidth]{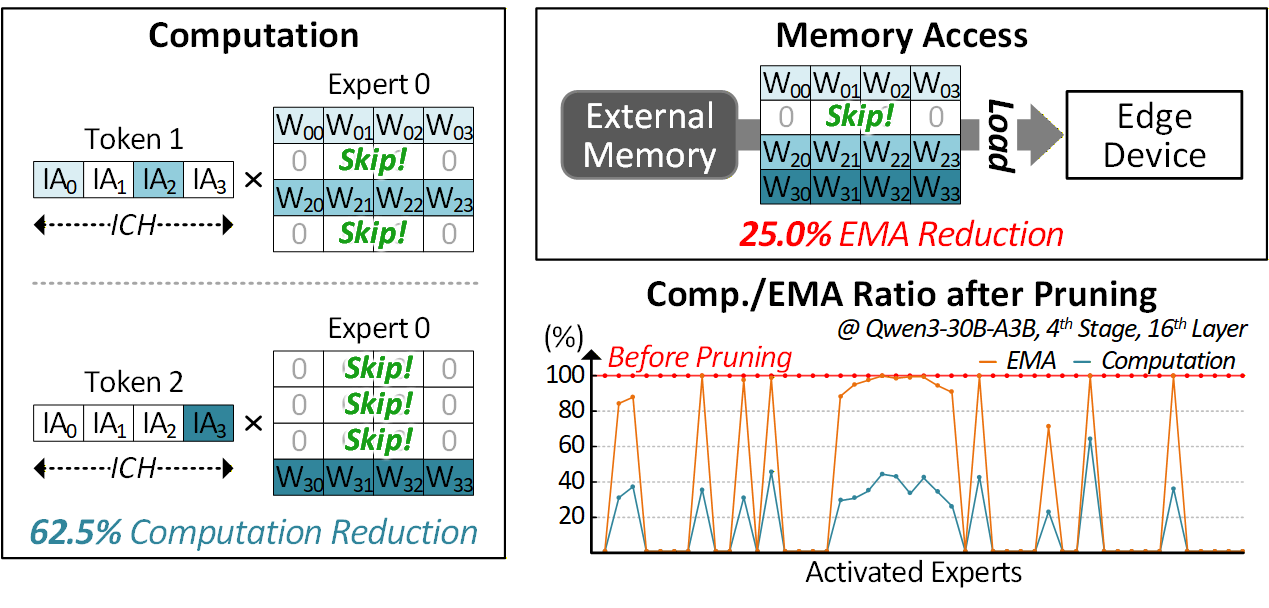}
    \caption{}\label{fig_9(a)}
  \end{subfigure}
  \hfill
  \begin{subfigure}[b]{\linewidth}
    \includegraphics[width=\linewidth]{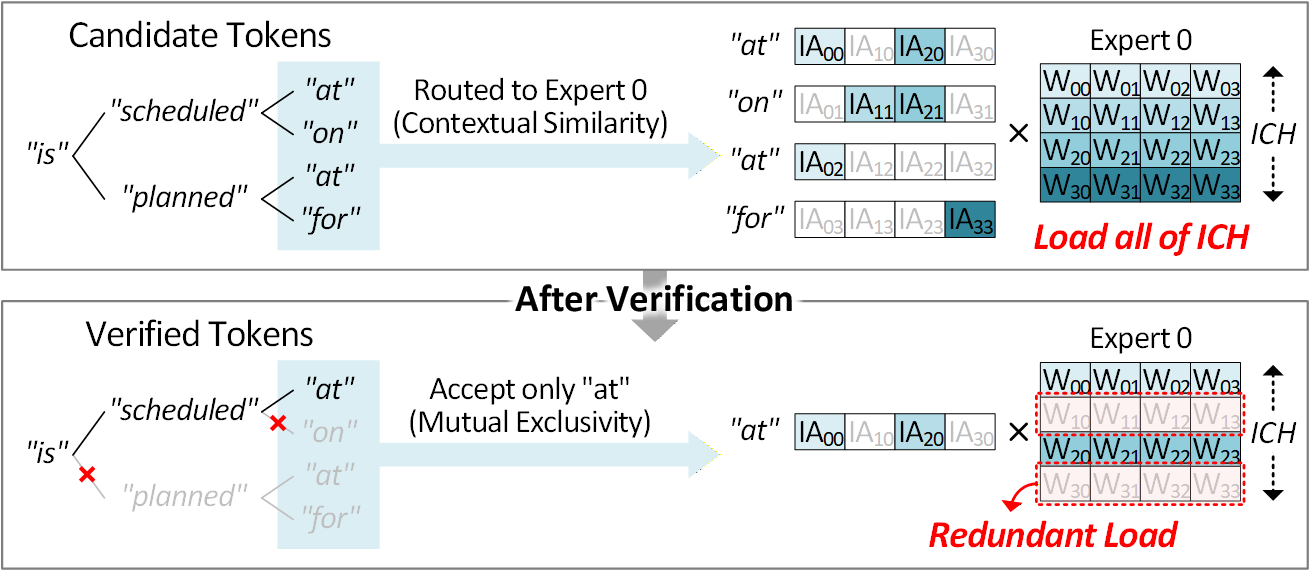}
    \caption{}\label{fig_9(b)}
  \end{subfigure}
  \hfill
  \begin{subfigure}[b]{\linewidth}
    \includegraphics[width=\linewidth]{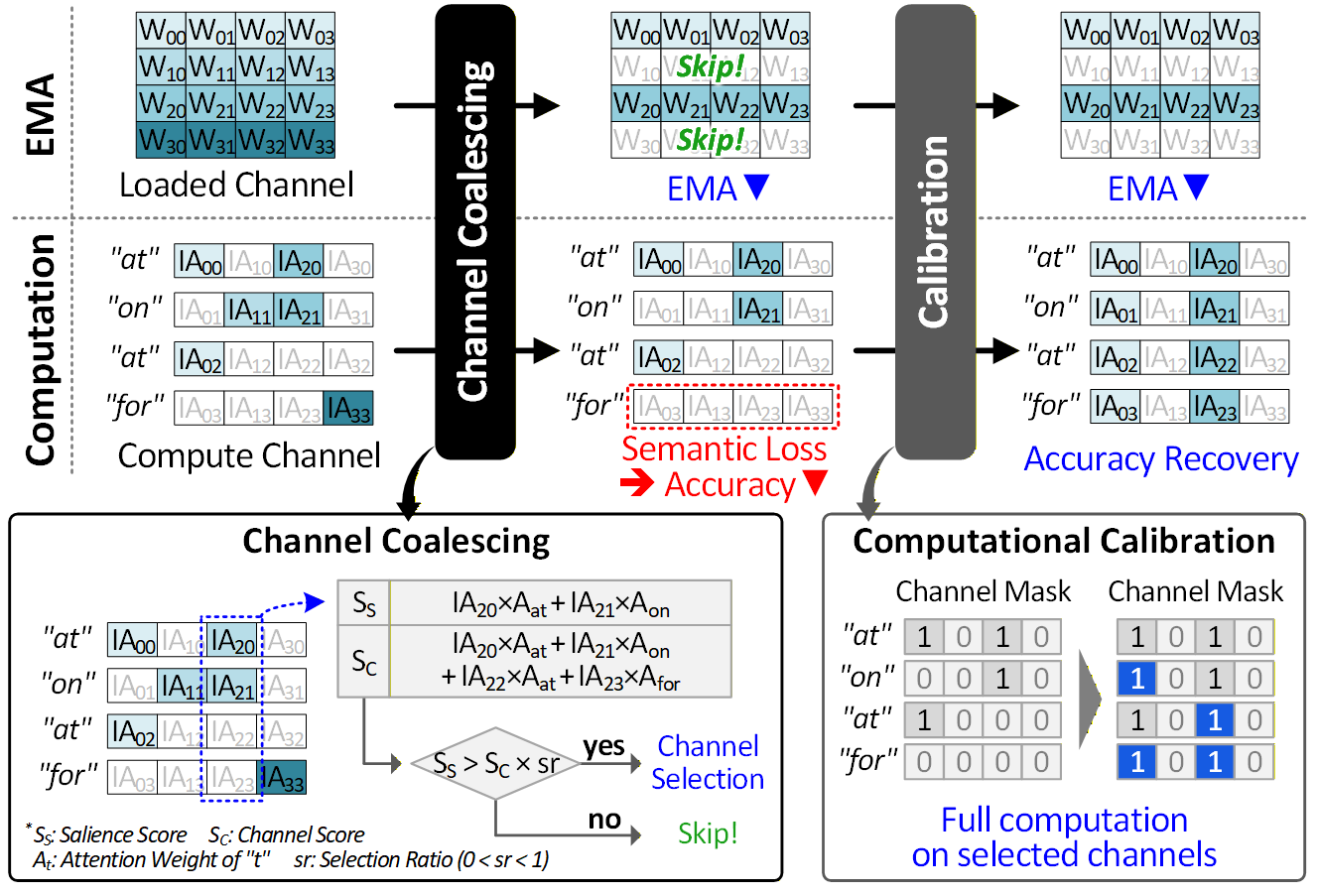}
    \caption{}\label{fig_9(c)}
  \end{subfigure}
  \hfill
  \begin{subfigure}[b]{\linewidth}
    \includegraphics[width=\linewidth]{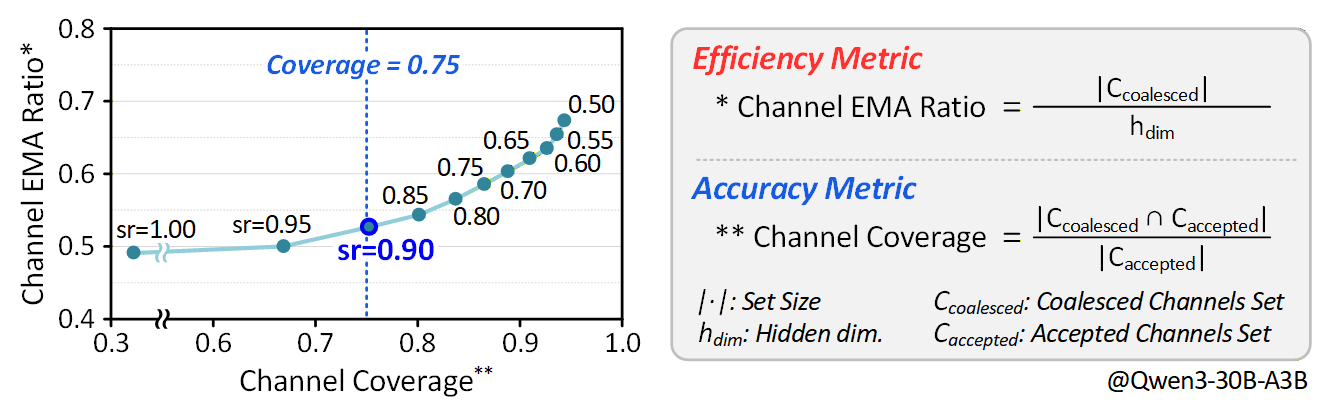}
    \vspace{-5mm}
    \caption{}\label{fig_9(d)}
  \end{subfigure}
  \hfill
  \captionsetup{justification=raggedright,singlelinecheck=false}
    \vspace{-4mm}
  \caption{(a) Limitation of pruning on EMA reduction.
  (b) Depth-wise redundancy.
  (c) Depth-aware expert coalescing.
  (d) Sensitivity of $sr$ to channel EMA and channel coverage.}
  \vspace{-5mm}
  \label{fig_9}
\end{figure}

While channel-level expert pruning \cite{smolpu, moe-pruner, e_prun2} has proven effective for MoE compression, its EMA reduction becomes limited when speculative decoding is applied. In speculative decoding, multiple candidate tokens are passed to the target MoE model. As shown in Figure~\ref{fig_9}(\subref{fig_9(a)}), when experts are pruned according to the channel requirements of individual candidate tokens, different tokens tend to select different channel subsets. Each token only computes on its own selected channel subset, enabling compute reduction to follow the pruning ratio. In contrast, since the hardware must load the union of all necessary channels to serve multiple tokens, EMA reduction remains marginal even when individual tokens utilize only sparse channel subsets. As the weight EMA is the primary bottleneck compared to computation, optimization is necessary. We propose depth-aware expert coalescing, a decode stage optimization that converts channel-level sparsity into actual EMA reduction.

EdgeXpert exploits two key characteristics of candidate tokens to reduce EMA during expert loading: depth-wise contextual similarity and mutual exclusivity. Depth-wise contextual similarity implies that tokens at the same depth exhibit similar semantic features and therefore tend to route to the same experts. Simultaneously, mutual exclusivity guarantees that only one token per depth is accepted at the verification stage, while all others are discarded. As illustrated in Figure~\ref{fig_9}(\subref{fig_9(b)}), when multiple same-depth tokens route to the same expert, the hardware must load the union of all required channels. Since verification discards all but one token per depth, most loaded channels serve discarded tokens, resulting in redundant EMA. Therefore, it is not necessary to retain the full union of all required channels. This allows EdgeXpert to apply channel coalescing within each depth with minor accuracy drop.

Figure~\ref{fig_9}(\subref{fig_9(c)}) presents our depth-aware expert coalescing. Instead of loading the union of all channels for tokens at the same depth, we selectively load only salient channels based on two metrics, the salience score ($S_s$) and the channel score ($S_c$).  The channel score aggregates the product of the input channel magnitude and attention weight ($|IA| \times A_t$) across all routed tokens, whereas the salience score considers only tokens that require computation on that channel, as determined by the channel pruning mask. Channels are selected for loading when their salience score exceeds $sr \times S_C$, where $sr$ is the selection ratio ($0<sr<1$), and others are skipped to reduce EMA. Although channel coalescing substantially reduces EMA, it may exclude channels that are semantically important for specific tokens. This semantic loss degrades accuracy and lowers the acceptance length, which in turn limits the overall EMA reduction.

To address this issue, we perform computational calibration whenever channel coalescing is applied. Calibration allows all routed tokens to recompute on all channels that have already been loaded by coalescing, rather than restricting each token to its own individually pruned subset. In this way, the loaded salient channels are more fully shared across tokens, which reduces the approximation error introduced by coalescing. Even if a channel important to one token is skipped, that token can still benefit from other salient channels that were loaded for different tokens at the same depth. While this increases computation, the trade-off remains favorable because it requires no additional memory access. Since EMA is the dominant bottleneck on edge devices, the extra computation helps recover both accuracy and acceptance length while maintaining the reduced EMA. Consequently, depth-aware expert coalescing reduces overall system energy and latency.

As shown in Figure~\ref{fig_9}(\subref{fig_9(d)}), we determine the selection ratio \(sr\) using the channel coverage of the accepted token after verification. For each depth group, we compare the coalesced channel set with the original pruned channel set of the accepted token. We select the maximum \(sr\) that maintains the average channel coverage above 75\%, ensuring that coalescing preserves channels required by the accepted token.

\begin{figure}
  \centering
  \begin{subfigure}[b]{\linewidth}
    \centering
    \includegraphics[width=0.9\linewidth]{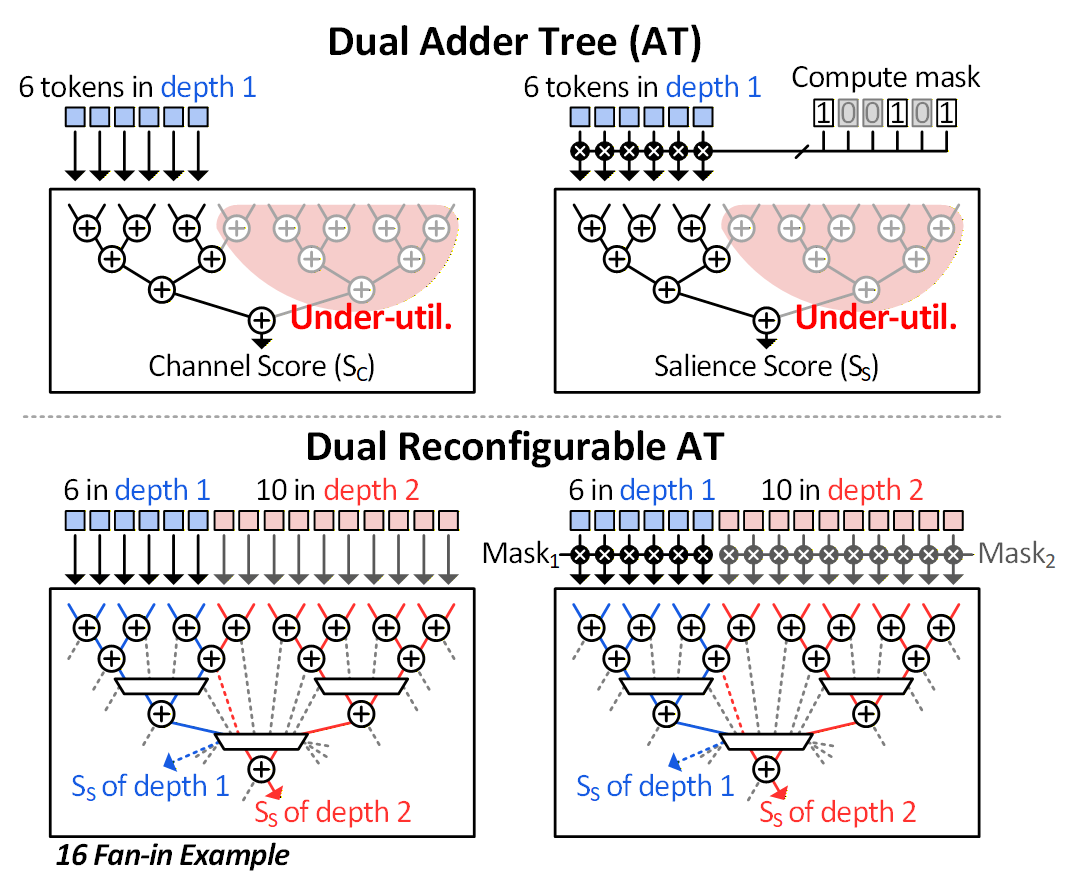}
    \vspace{-2mm}
    \caption{}\label{fig_10(a)}
  \end{subfigure}
  \hfill
  \begin{subfigure}[b]{\linewidth}
    \includegraphics[width=\linewidth]{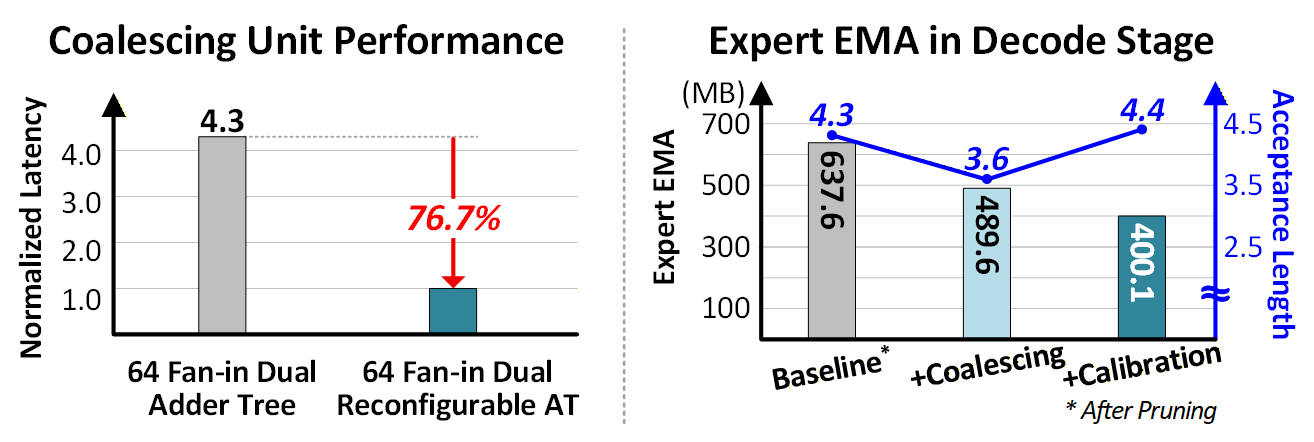}
    \caption{}\label{fig_10(b)}
  \end{subfigure}
  \hfill
  \captionsetup{justification=raggedright,singlelinecheck=false}
    \vspace{-3mm}
  \caption{(a) Dual reconfigurable adder tree in coalescing unit.
  (b) Performance analysis.}
    \vspace{-2mm}
  \label{fig_10}
\end{figure}

Figure~\ref{fig_10}(\subref{fig_10(a)}) shows the dual reconfigurable adder tree used to compute the channel and salience scores for channel coalescing. Because the number of candidate tokens varies across depths, a conventional fixed adder tree suffers from severe under-utilization. We therefore adopt a reconfigurable adder tree (RAT) \cite{sigma}, which can process multiple depth groups within a single unit by accumulating same-depth inputs while bypassing others. This allows a RAT to handle multiple depths at once and reduces coalescing latency.

Figure~\ref{fig_10}(\subref{fig_10(b)}) shows that the dual RAT reduces coalescing unit latency by 76.7\% compared with a conventional dual adder tree. Although the RAT incurs additional area and power compared to a fixed adder tree, the impact is minor because the top controller accounts for only a small fraction of the total area and power. Figure~\ref{fig_10}(\subref{fig_10(b)}) also shows the effect of depth-aware expert coalescing on expert EMA. The baseline applies only coarse-to-fine pruning \cite{smolpu} without further optimization. Channel coalescing reduces the amount of loaded expert data, but also lowers the acceptance length, limiting the EMA reduction. When computational calibration is applied after each coalescing step, the acceptance length is recovered. This reduces the number of decoding stages, which further reduces expert EMA. These results show that EdgeXpert converts channel sparsity into actual EMA reduction while preserving speculative decoding effectiveness. This makes MoE and speculative decoding more suitable for edge deployment.

\section{Evaluation}
\label{sec:evaluation}
\subsection{Experimental Setup}

EdgeXpert was synthesized using Synopsys Design Compiler, targeting Samsung 28nm FDSOI technology at typical corner conditions (TT, 25°C, 1.0V). The design flow consists of RTL design, synthesis, logic optimization, post-synthesis timing check, and power analysis using Synopsys PrimeTime. The synthesized logic area is projected to the post-PnR area, assuming 60\% cell utilization, a typical utilization rate for digital designs to accommodate routing congestion and timing closure. SRAM macros are modeled using a commercial memory compiler for the target technology node. The total chip area is calculated as the sum of the projected logic area (synthesized area divided by 0.6) and the SRAM macro area.

EdgeXpert operates at 800 MHz with a supply voltage of 1.0V. The hardware supports 8-bit input activation and 4-bit weight (A8W4) for efficient edge inference. This configuration delivers a peak throughput of 13.1 TOPS, calculated as 8K MACs multiplied by 800 MHz and 2 operations per MAC. Hardware power is obtained from post-synthesis power analysis using Synopsys PrimeTime. External memory consists of LPDDR4X DRAM with 16 GB/s bandwidth. We estimate LPDDR4X DRAM read and write energies from the datasheet IDD currents and supply voltages of the LPDDR4X power rails~\cite{micron_dram}, resulting in 4.5 pJ/bit and 2.9 pJ/bit, respectively. Physical design metrics show a synthesized logic area of 3.8 mm$^2$ before PnR, projected to 6.3 mm$^2$ assuming 60\% cell utilization. SRAM macros occupy 3.0 mm$^2$. The total chip area, including logic and SRAM, measures 9.3 mm$^2$. Hardware peak power is 1.52 W without EMA power. The hardware achieves a peak energy efficiency of 8.6 TOPS/W without sparsity exploitation.

EdgeXpert is evaluated using an instruction-level custom simulator. The simulator implements behavioral models of the proposed hardware modules, including router reformulation, channel coalescing, and computational calibration. Given the original model trace, including expert routing information and the token tree, the behavioral models emulate the hardware-module behavior to derive the final expert/channel access pattern. The simulator then generates the corresponding runtime instruction sequence and memory requests, and accumulates latency and energy using per-instruction costs. The per-instruction costs for logic and SRAM are obtained from post-synthesis results, while the EMA cost is computed from the read/write traffic. For off-chip memory, the simulator generates read/write requests from the final expert/channel access pattern and computes their latency using the 16 GB/s bandwidth. The energy is accumulated from the read/write volume using the energy parameters in the datasheet~\cite{micron_dram}.


\begin{table}[!t]
\centering
\caption{Hardware configuration for comparison}
\label{table_1}
\footnotesize
\setlength{\tabcolsep}{6pt}
\renewcommand{\arraystretch}{1.15}
\begin{tabular}{@{}c|c|c|c|c@{}}
\toprule
 & \textbf{\shortstack{Ext. BW}} & \textbf{\shortstack{DRAM}}
 & \textbf{\shortstack{Throughput}} & \textbf{\shortstack{Peak Power}} \\
\midrule
\textbf{MoE-Pruner$_{\rm AU}$} & 16 GB/s & LPDDR4X & 13.1 TOPS & 2.49$^{*}$ W \\
\textbf{MoE-Pruner$_{\rm SD}$} & 16 GB/s & LPDDR4X & 13.1 TOPS & 2.49$^{*}$ W \\
\textbf{EdgeMoE$_{\rm AU}$} & 16 GB/s & LPDDR4X & 13.1 TOPS & 1.29$^{**}$ W \\
\textbf{EdgeMoE$_{\rm SD}$} & 16 GB/s & LPDDR4X & 13.1 TOPS & 1.29$^{**}$ W \\
\textbf{SMoLPU} & 16 GB/s & LPDDR4X & 14.4 TOPS & 1.44 W \\
\midrule[0.7pt]
\textbf{EdgeXpert} & 16 GB/s & LPDDR4X & 13.1 TOPS & 1.52 W \\
\bottomrule
\end{tabular}
 
\vspace{2pt}
{\footnotesize\raggedleft
$^{*}$~Arbitrary skipping MAC~\cite{space-mate} \quad
$^{**}$~Dense MAC~\cite{scnn}\par}
\end{table}

Table~\ref{table_1} summarizes the hardware configurations. We compare EdgeXpert against two hardware and one software baseline.

\textit{\textbf{Hardware Baseline.}} SMoLPU \cite{smolpu} is a dedicated accelerator for MoE-based speculative decoding that applies coarse-to-fine expert pruning during the decode stage. EdgeMoE \cite{edgemoe} reduces expert footprint through importance-aware mixed-precision quantization, assigning 4-bit or 2-bit precision to each expert. For fair comparison, we project EdgeMoE onto dedicated hardware with 4-bit dense INT MAC units \cite{scnn} rather than comparing directly on the Jetson platform. Since it originally targets autoregressive decoding, we evaluate it under both autoregressive decoding (EdgeMoE\textsubscript{AU}) and speculative decoding (EdgeMoE\textsubscript{SD}).

\textit{\textbf{Software Baseline.}} Due to the limited number of existing MoE hardware accelerators, we additionally compare against MoE-Pruner \cite{moe-pruner}, a state-of-the-art software pruning scheme that applies fine-grained unstructured sparsity within each expert. We map MoE-Pruner onto a hardware accelerator that supports arbitrary skipping \cite{space-mate}. Since MoE-Pruner targets autoregressive decoding, we evaluate both autoregressive (MoE-Pruner\textsubscript{AU}) and speculative decoding (MoE-Pruner\textsubscript{SD}). Although MoE-Pruner supports optional fine-tuning after pruning, we disable it to ensure consistency.


While Table~\ref{table_0} summarizes related speculative decoding works for MoE models, MoESD~\cite{moesd} and SS-MoE~\cite{ss-moe} are not directly comparable to our edge accelerator setting. MoESD focuses on GPU-level operating-point analysis, while SS-MoE adopts a large self-draft model, which is less suitable for edge devices. MoE-Spec~\cite{moe-spec} performs only coarse expert budgeting during verification. Therefore, SMoLPU, which supports coarse-to-fine pruning, provides a stronger pruning baseline. Accordingly, our quantitative comparison covers three representative categories: fine-grained pruning for MoE models (MoE-Pruner), mixed-precision quantization for MoE models (EdgeMoE), and coarse-to-fine pruning for MoE-based speculative decoding (SMoLPU).

All baselines are normalized to the same technology (28nm CMOS), operating frequency (800 MHz), and external DRAM bandwidth (16 GB/s) as EdgeXpert. SMoLPU is scaled from its original 200 MHz to 800 MHz, isolating architectural efficiency from clock-frequency differences. Since EdgeMoE originally reports power from a Jetson board, we map its mixed-precision execution to the EdgeXpert backend with 4-bit dense INT MAC units. This preserves the benefits of mixed precision while excluding the board-level system power. MoE-Pruner is mapped to hardware with arbitrary skipping cores so that its unstructured sparsity directly reduces latency and energy. All evaluations perform single-batch inference, reflecting edge deployment scenarios.



\begin{table}[!t]
\centering
\caption{Model configuration}
\label{table_2}
\footnotesize
\setlength{\tabcolsep}{3pt}
\renewcommand{\arraystretch}{1.1}
\begin{tabular}{@{}c|c|c|c|c@{}}
\toprule
 & \textbf{Granite} & \textbf{OLMoE} & \textbf{DeepSeek} & \textbf{Qwen3} \\
 & \textbf{1B-A400M} & \textbf{1B-7B} & \textbf{V2-Lite} & \textbf{30B-A3B} \\
\midrule[0.7pt]
\multicolumn{5}{c}{\textbf{Target Model}} \\
\midrule[0.7pt]
\# Layer        & 24   & 16   & 27 (26 MoE) & 48 \\
Hidden Dim.     & 1024 & 2048 & 2048 & 2048 \\
\# Experts      & 32   & 64   & 64   & 128 \\
Expert Param.   & 1.5M & 6.0M & 8.3M & 4.5M \\
Top-$k$ Routing & 8    & 8    & 8 (2 shared exp.) & 8 \\
\midrule[0.7pt]
\multicolumn{5}{c}{\textbf{Draft Model}} \\
\midrule[0.7pt]
\# Layer    & 1    & 1    & 1    & 1 \\
Hidden Dim. & 1024 & 2048 & 2048 & 2048 \\
Max Depth   & 6    & 6    & 6    & 6 \\
\midrule[0.7pt]
\multicolumn{5}{c}{\textbf{Hyperparameters}} \\
\midrule[0.7pt]
$k_{min}, n_{min}$       & 8, 7  & 10, 6 & 6, 5 & 10, 6 \\
$k_{tri}, n_{tri}$       & 10, 6 & 12, 4 & 8, 3 & 12, 4 \\
$\theta_1, \theta_2, sr$ & .65, .45, .70 & .80, .40, .90 & .55, .30, .90 & .80, .55, .90 \\
\bottomrule
\end{tabular}

\vspace{2pt}
\end{table}



\begin{table*}[!t]
\centering
\caption{Accuracy and acceptance length}
\label{table_3}
\footnotesize
\begin{tabular*}{\textwidth}{@{\extracolsep{\fill}}c|c|ccccccc|c@{}}
\toprule
\textbf{Model} & \textbf{Hardware} & \textbf{MT-Bench} & \textbf{GSM8K} & \textbf{MMLU}
 & \textbf{HellaSwag} & \textbf{ARC-C} & \textbf{WinoGrande} & \textbf{PIQA}
 & \textbf{Acc. Len.}$^{*}$ \\
\midrule
\multirow{4}{*}{Granite-1B-A400M}
 & Baseline & 5.3 & 29.3\% & 29.6\% & 58.4\% & 33.5\% & 58.0\% & 70.5\% & 3.9 \\
 & EdgeXpert$_{\rm PR}$ & 5.2 & 27.2\% & 27.5\% & 56.1\% & 33.0\% & 55.7\% & 69.9\% & 3.9 \\
 & EdgeXpert$_{\rm DC}$ & 2.8 & 17.1\% & 27.5\% & 56.1\% & 33.0\% & 55.7\% & 69.9\% & 3.6 \\
 & \textbf{EdgeXpert$_{\rm ALL}$} & \textbf{5.2} & \textbf{27.7\%} & \textbf{27.5\%}
 & \textbf{56.1\%} & \textbf{33.0\%} & \textbf{55.7\%} & \textbf{69.9\%} & \textbf{3.9} \\
\midrule
\multirow{4}{*}{OLMoE-1B-7B}
 & Baseline & 6.2 & 62.6\% & 51.4\% & 71.1\% & 45.8\% & 63.9\% & 76.2\% & 4.0 \\
 & EdgeXpert$_{\rm PR}$ & 6.1 & 62.6\% & 49.1\% & 68.6\% & 43.9\% & 64.9\% & 75.8\% & 4.0 \\
 & EdgeXpert$_{\rm DC}$ & 4.2 & 16.5\% & 49.1\% & 68.6\% & 43.9\% & 64.9\% & 75.8\% & 2.7 \\
 & \textbf{EdgeXpert$_{\rm ALL}$} & \textbf{6.2} & \textbf{58.9\%} & \textbf{49.1\%}
 & \textbf{68.6\%} & \textbf{43.9\%} & \textbf{64.9\%} & \textbf{75.8\%} & \textbf{4.1} \\
\midrule
\multirow{4}{*}{DeepSeek-V2-Lite}
 & Baseline & 6.5 & 60.5\% & 55.3\% & 72.4\% & 46.1\% & 68.2\% & 78.6\% & 3.8 \\
 & EdgeXpert$_{\rm PR}$ & 6.5 & 60.2\% & 55.4\% & 72.2\% & 45.1\% & 67.2\% & 78.2\% & 3.8 \\
 & EdgeXpert$_{\rm DC}$ & 5.4 & 15.5\% & 55.4\% & 72.2\% & 45.1\% & 67.2\% & 78.2\% & 3.2 \\
 & \textbf{EdgeXpert$_{\rm ALL}$} & \textbf{6.4} & \textbf{59.7\%} & \textbf{55.4\%}
 & \textbf{72.2\%} & \textbf{45.1\%} & \textbf{67.2\%} & \textbf{78.2\%} & \textbf{4.0} \\
\midrule
\multirow{4}{*}{Qwen3-30B-A3B}
 & Baseline & 8.8 & 72.8\% (87.3\%$^{**}$) & 76.2\% & 75.0\% & 50.2\% & 70.3\% & 79.2\% & 4.3 \\
 & EdgeXpert$_{\rm PR}$ & 8.7 & 72.8\% (87.3\%$^{**}$) & 74.4\% & 77.1\% & 49.6\% & 68.1\% & 76.1\% & 4.3 \\
 & EdgeXpert$_{\rm DC}$ & 5.8 & 66.8\% (76.1\%$^{**}$) & 74.4\% & 77.1\% & 49.6\% & 68.1\% & 76.1\% & 3.6 \\
 & \textbf{EdgeXpert$_{\rm ALL}$} & \textbf{8.7} & \textbf{69.4\% (88.0\%$^{**}$)} & \textbf{74.4\%}
 & \textbf{77.1\%} & \textbf{49.6\%} & \textbf{68.1\%} & \textbf{76.1\%} & \textbf{4.4} \\
\bottomrule
\end{tabular*}
 
\vspace{2pt}
{\footnotesize\raggedleft
$^{*}$~Acceptance length\quad
$^{**}$~Thinking mode (multi-step reasoning)\par}
\end{table*}

\begin{figure}[t]
  \centering
    \hspace*{\fill} 
    \includegraphics[width=0.95\linewidth]{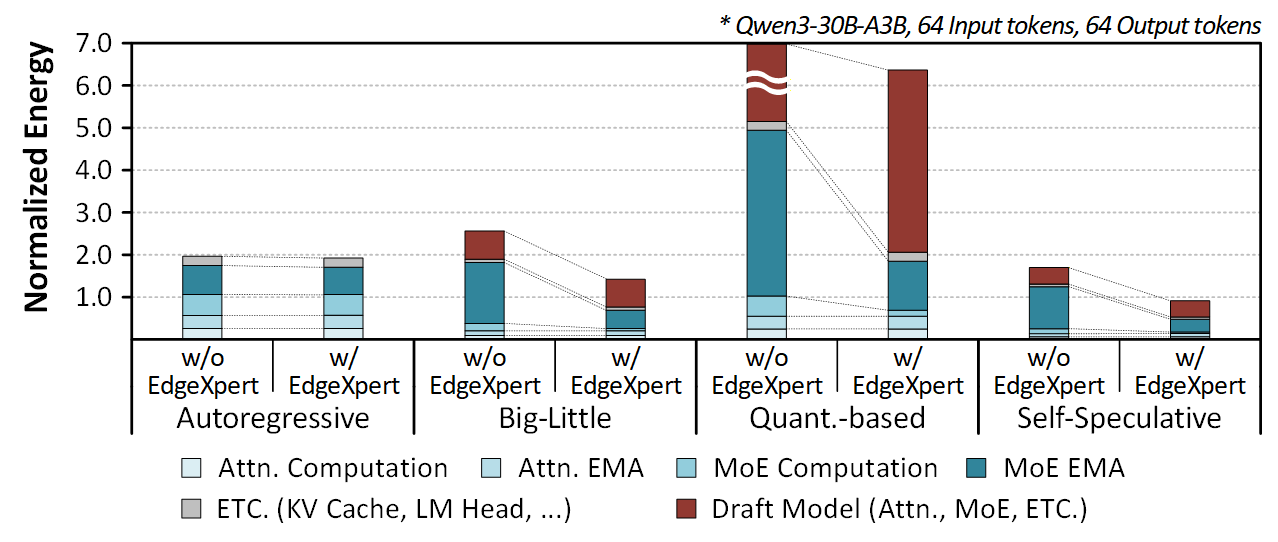}
    \hspace*{\fill} 
  \caption{Comparison of speculative decoding frameworks.}
  \label{fig_11}
\end{figure}

As shown in Table~\ref{table_2}, we evaluated four representative MoE models \cite{granite, olmoe, deepseek-v2, qwen3}. Since EdgeXpert optimizes MoE models during the verification phase, it is applicable to various speculative decoding frameworks. Figure~\ref{fig_11} compares the energy consumption of different frameworks. Self-speculative decoding achieves the lowest energy consumption because its draft model is lightweight while still providing a high acceptance length. Based on this result, we adopt EAGLE-3 \cite{eagle3}, a self-speculative decoding framework that generates 64 candidate tokens per iteration with a maximum depth of 6. Moreover, we evaluated accuracy on seven benchmarks \cite{mt-bench, gsm8k, mmlu, hellaswag, arc_challenge, winogrande, piqa}. For the hyperparameters $k_{min,tri}$, $n_{min,tri}$, and $sr$, we do not tune them on downstream benchmark accuracy. Instead, we determine them using model-distribution-based metrics, i.e., MoE-layer output deviation and accepted-token channel coverage. This design reduces benchmark-specific sensitivity, since the hyperparameters are selected to preserve the original model behavior rather than to overfit a particular evaluation task. As a result, the selected values maintain accuracy across benchmarks. The calibration is performed only once per model: \(k_{min,tri}\) and \(n_{min,tri}\) are calibrated on WikiText-2 validation set~\cite{wikitext}, while \(sr\) is selected using a small calibration set sampled from MT-Bench prompts to preserve the original channel distribution, rather than to optimize benchmark accuracy. For the partitioning thresholds $\theta_{1,2}$, we use WikiText-2 perplexity. Although the encoder is shared across MoE models, each MoE model exhibits different sensitivity to budgeted routing. Therefore, $\theta_{1,2}$ is calibrated once for each model.

\subsection{Experimental Results}
\textit{\textbf{Accuracy and acceptance length analysis.}} Table~\ref{table_3} presents the accuracy and acceptance length of the proposed techniques. We evaluate four configurations by incrementally enabling each technique. Baseline combines MoE and speculative decoding with the pruning scheme of prior work~\cite{smolpu}. EdgeXpert\textsubscript{PR} applies prompt-wise expert reuse on top of Baseline. EdgeXpert\textsubscript{DC} further introduces depth-aware channel coalescing. EdgeXpert\textsubscript{ALL} incorporates all proposed techniques, including computational calibration. Prompt-wise expert reuse introduces negligible accuracy loss and leaves the acceptance length unchanged. In particular, its effect in the prefill stage can be observed on MMLU, HellaSwag, ARC-Challenge, WinoGrande, and PIQA, where the maximum degradation is only 1.0-3.1\% across the evaluated models. This indicates that major tokens effectively establish the shared expert set in the prefill stage, while the remaining tokens can largely be served through reuse without significant quality loss. Depth-aware channel coalescing reduces EMA, but significantly degrades both accuracy and acceptance length. However, with computational calibration, both metrics are recovered to near-baseline levels. The final drop in MT-Bench score remains within 0.1, while GSM8K shows a larger degradation of 3--4\%. This degradation is mainly due to the sensitivity of GSM8K under the non-thinking inference. The non-thinking setting provides fewer intermediate reasoning steps, leaving fewer chances to recover from such approximation errors introduced by channel coalescing. In contrast, the longer reasoning trace in thinking mode mitigates this sensitivity and eliminates the additional accuracy drop. With Qwen3 in thinking mode (Table~\ref{table_3}), the proposed techniques introduce no accuracy drop on GSM8K.

\begin{figure}
  \centering
  \begin{subfigure}[b]{\linewidth}
    \includegraphics[width=\linewidth]{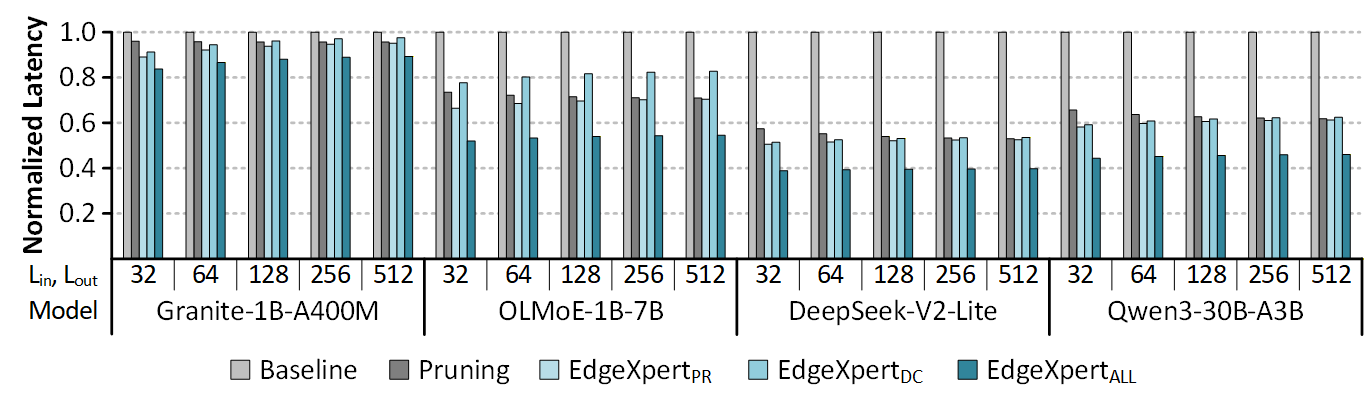}
    \caption{}\label{fig_12(a)}
  \end{subfigure}
  \hfill
  \begin{subfigure}[b]{\linewidth}
    \includegraphics[width=\linewidth]{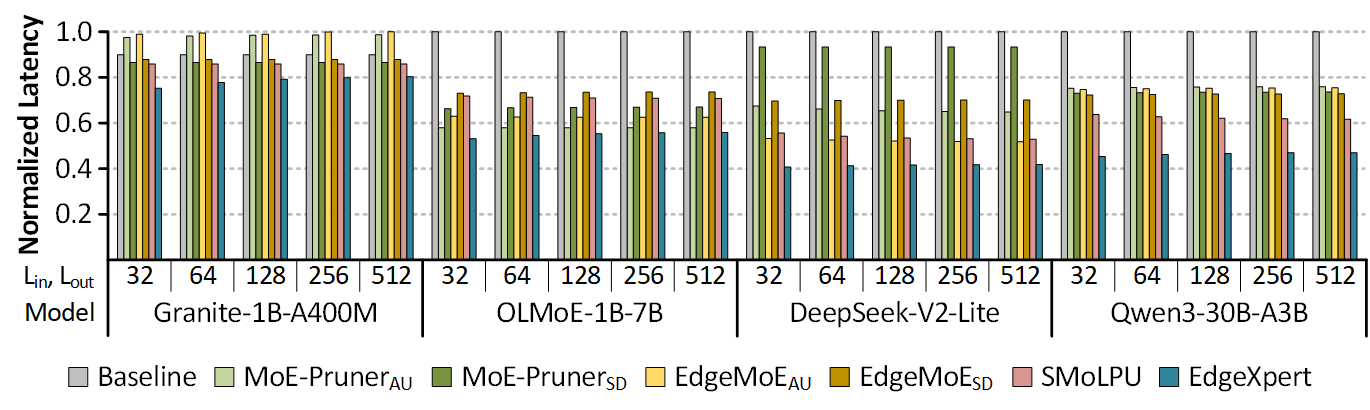}
    \caption{}\label{fig_12(b)}
  \end{subfigure}
  \hfill
  \begin{subfigure}[b]{\linewidth}
    \includegraphics[width=\linewidth]{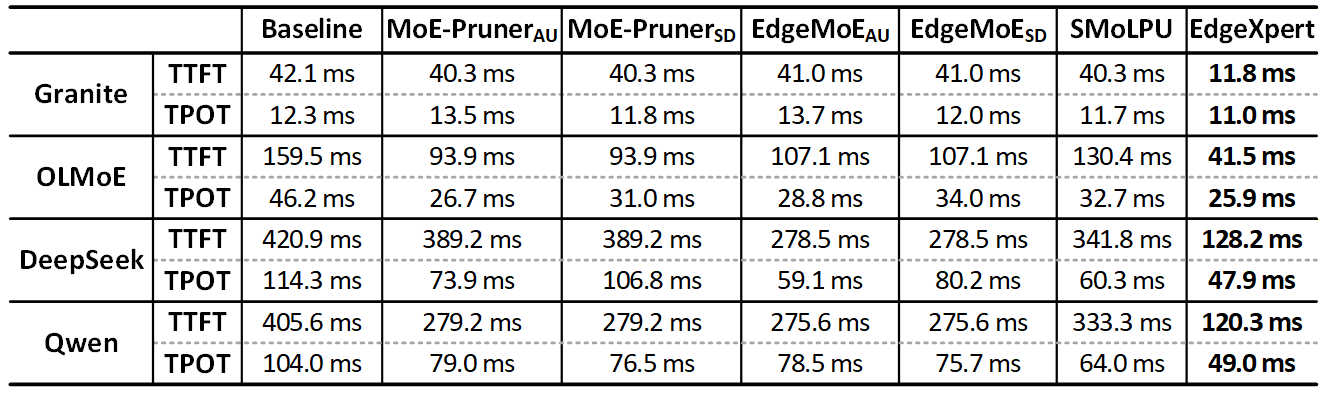}
    \caption{}\label{fig_12(c)}
  \end{subfigure}
  \hfill
  \captionsetup{justification=raggedright,singlelinecheck=false}
  \vspace{-4mm}
  \caption{(a) Normalized end-to-end latency of EdgeXpert.
  (b) Comparison of end-to-end latency against prior works.
  (c) Detailed TTFT and TPOT.}
  \vspace{-3mm}
  \label{fig_12}
\end{figure}

\textit{\textbf{Latency analysis.}} Figure~\ref{fig_12}(\subref{fig_12(a)}) shows the normalized end-to-end latency of EdgeXpert on multiple benchmarks. The baseline represents MoE combined with speculative decoding without any optimization. Pruning applies the coarse-to-fine expert pruning of previous work \cite{smolpu}. EdgeXpert\textsubscript{PR} reduces the latency of the prefill stage through prompt-wise expert reuse. The gain is larger at shorter token lengths, where the prefill stage constitutes a larger fraction of the total latency. EdgeXpert\textsubscript{DC} applies depth-aware channel coalescing on top of EdgeXpert\textsubscript{PR} and further reduces EMA. However, channel coalescing also lowers the acceptance length and limits the latency improvement. Applying computational calibration recovers the acceptance length. As shown in EdgeXpert\textsubscript{ALL}, this recovery translates into a large overall latency reduction. DeepSeek shows the largest latency reduction because its expert size is larger than that of the other evaluated models. EdgeXpert achieves 10.7-61.1\% latency reduction. Figure~\ref{fig_12}(\subref{fig_12(b)}) compares end-to-end latency with previous work. Baseline maps MoE with speculative decoding onto dense MAC \cite{scnn} without any optimization. Since MoE-Pruner \cite{moe-pruner} statically prunes expert weights, it can reduce latency in both autoregressive and speculative decoding. However, because the static pruning ratio is limited, the overall latency reduction remains modest. 
EdgeMoE \cite{edgemoe} assigns mixed precision to each expert based on importance at the offline. This reduces expert EMA at runtime and is also applicable to both autoregressive and speculative decoding (EdgeMoE\textsubscript{AU} and EdgeMoE\textsubscript{SD}). However, mixed-precision quantization reduces expert EMA by only 30–35\%, leading to limited latency improvement. SMoLPU \cite{smolpu} performs runtime pruning of redundant experts in a system that combines MoE and speculative decoding. It achieves the highest pruning ratio among the three prior works and therefore provides a larger latency reduction. EdgeXpert goes beyond pruning by introducing EMA-aware expert routing and loading, which further reduces EMA. Consequently, it achieves the lowest latency across all models, delivering up to 56.3\% end-to-end latency reduction compared to prior works.

\begin{figure}
  \centering
  \begin{subfigure}[b]{\linewidth}
    \includegraphics[width=\linewidth]{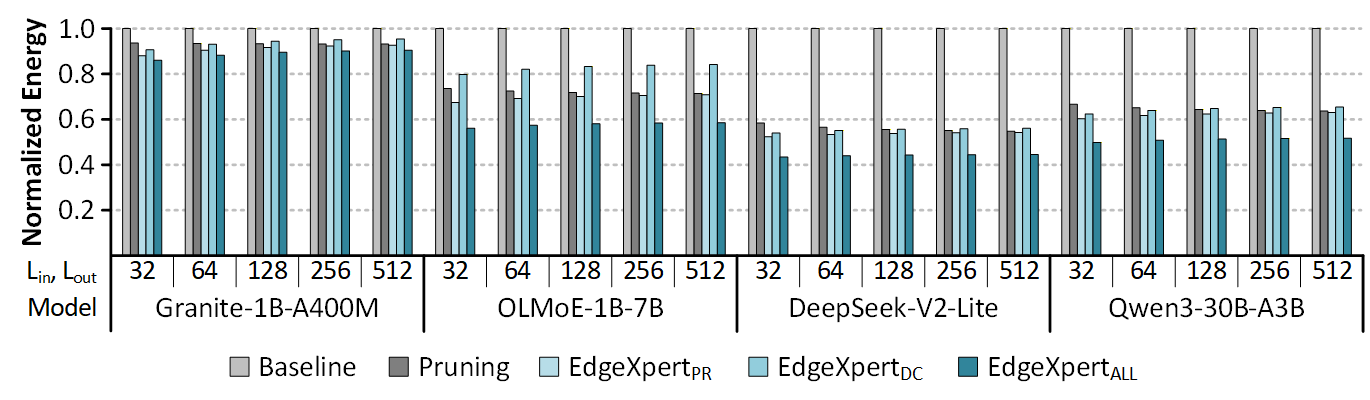}
    \caption{}\label{fig_13(a)}
  \end{subfigure}
  \hfill
  \begin{subfigure}[b]{\linewidth}
    \includegraphics[width=\linewidth]{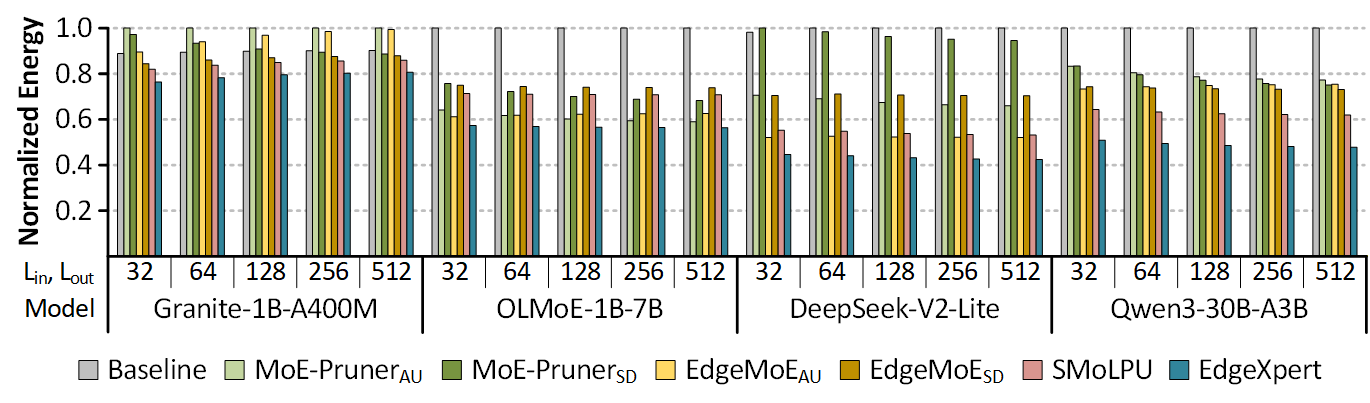}
    \caption{}\label{fig_13(b)}
  \end{subfigure}
  \hfill
  \begin{subfigure}[b]{\linewidth}
    \includegraphics[width=\linewidth]{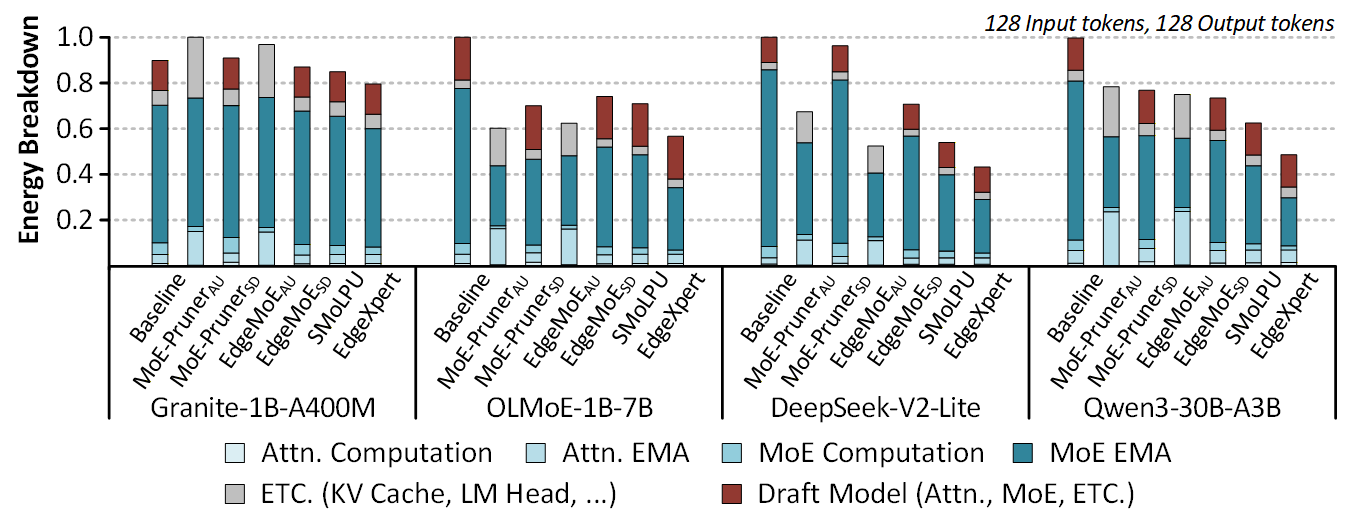}
    \caption{}\label{fig_13(c)}
  \end{subfigure}
  \hfill
  \captionsetup{justification=raggedright,singlelinecheck=false}
  \vspace{-4mm}
  \caption{(a) Normalized energy consumption of EdgeXpert.
  (b) Comparison of energy consumption against prior works.
  (c) Detailed energy breakdown.}
  \label{fig_13}
\end{figure}

Figure~\ref{fig_12}(\subref{fig_12(c)}) presents the detailed time to first token (TTFT) and time per output token (TPOT). Practical edge deployment requires TTFT below 450 ms and TPOT below 50 ms \cite{ttft_tpot}. Prompt-wise expert reuse effectively reduces TTFT, where the proposed reuse policy lowers expert EMA in the prefill. Moreover, depth-aware expert coalescing reduces TPOT, where selective channel loading reduces expert EMA. As a result, EdgeXpert is the only design that satisfies both thresholds across all evaluated benchmarks.

\textit{\textbf{Energy consumption analysis.}} Figure~\ref{fig_13}(\subref{fig_13(a)}) shows the normalized energy consumption of EdgeXpert. Baseline combines MoE with speculative decoding without any optimization, while pruning applies coarse-to-fine pruning. Similar to the latency trend, EdgeXpert\textsubscript{PR} provides larger energy reduction at shorter token lengths. EdgeXpert\textsubscript{ALL} achieves the largest overall energy reduction (9.5-56.6\%) thanks to EMA-aware expert routing and loading. Figure~\ref{fig_13}(\subref{fig_13(b)}) compares the energy consumption with prior works. Although the energy reduction ratio varies depending on the pruning ratio of each model, EdgeXpert consistently achieves the lowest energy consumption across all models, delivering up to 44.1\% energy reduction compared to prior works. Figure~\ref{fig_13}(\subref{fig_13(c)}) shows the energy breakdown of each design. Baseline applies speculative decoding to MoE without optimization. In MoE-Pruner and EdgeMoE, applying speculative decoding increases both draft model overhead and MoE layer overhead compared to autoregressive decoding. However, the energy reduction from expert pruning and quantization is larger under speculative decoding because EMA of the MoE layer accounts for more of the total system energy. Speculative decoding also reduces the energy consumption of the attention layer and other components, such as the LM head and KV cache. EdgeXpert further reduces the EMA and computation overhead of the MoE layers, achieving the lowest overall energy consumption.


\begin{table}[!t]
\centering
\caption{Synthesis results of EdgeXpert}
\label{table_4}
\resizebox{\columnwidth}{!}{%
\footnotesize
\setlength{\tabcolsep}{2pt}
\renewcommand{\arraystretch}{1.15}
\begin{tabular}{@{}c|c|c|c|c@{}}
\toprule
 & & \textbf{Components} & \textbf{Area [mm$^{2}$]} & \textbf{Power [W]} \\
\midrule[0.7pt]
\multirow{3}{*}{\rotatebox[origin=c]{90}{\textbf{\shortstack{Cluster\\\#0\ --\ \#7}}}}
 & \textbf{PE Lines}   & 8$\times$4$\times$8$\times$32 PE & 2.24 (24.1\%) & 0.70 (46.1\%) \\
 & \textbf{SRAM}       & WMEM, IAMEM, SF$_{\rm W}$MEM     & 1.70 (18.3\%) & 0.19 (12.5\%) \\
 & \textbf{Peripheral} & Workload allocator, \ldots       & 1.27 (13.6\%) & 0.12 (7.9\%)  \\
\midrule[0.7pt]
\multirow{7}{*}{\rotatebox[origin=c]{90}{\textbf{\shortstack{Top\\Controller}}}}
 & \multirow{3}{*}{\textbf{PRU}$^{*}$}
   & Partitioning network       & 0.11 (1.2\%) & $<$0.01 (0.4\%) \\
 & & Expert union unit          & 0.16 (1.7\%) & $<$0.01 (0.2\%) \\
 & & Etc. (Routing map, \ldots) & 0.04 (0.4\%) & $<$0.01 (0.1\%) \\
\cline{2-5}
 & \multirow{3}{*}{\textbf{DCU}$^{**}$}
   & Channel coalescing unit    & 0.26 (2.8\%) & 0.03 (2.1\%) \\
 & & Compute mask update unit   & 0.05 (0.5\%) & 0.01 (0.5\%) \\
 & & Etc. (Channel REG, \ldots) & 0.15 (1.6\%) & 0.01 (0.7\%) \\
\cline{2-5}
 & \textbf{Etc.} & Pruning logic, Addr. generator & 0.55 (5.9\%) & 0.07 (4.6\%) \\
\midrule[0.7pt]
\multicolumn{2}{c|}{\textbf{Shared SRAM}} & 512KB SRAM
 & 1.34 (14.4\%) & 0.21 (13.8\%) \\
\midrule[0.7pt]
\multicolumn{2}{c|}{\textbf{Others}} & SIMD, NoC, Instruction Dec.
 & 1.44 (15.5\%) & 0.17 (11.1\%) \\
\bottomrule
\end{tabular}}
 
\vspace{2pt}
{\scriptsize\raggedleft
$^{*}$Prompt-wise Expert Reuse Unit \quad
$^{**}$Depth-aware Coalescing Unit\par}
\end{table}

\textit{\textbf{Area and power overhead.}} Table~\ref{table_4} presents the area and average power of EdgeXpert. The top controller, which consists of a prompt-wise expert reuse unit (PRU) and a depth-aware coalescing unit (DCU), occupies only 14.1\% of the total chip area and 8.6\% of total power due to two factors. First, prompt-wise expert reuse is executed once for each layer, and depth-aware expert coalescing is applied once for each expert, whereas the clusters repeatedly perform MAC operations. Second, controller execution overlaps with long-latency cluster MAC operations, amortizing its power across extended computation periods. Moreover, the latency overhead of the top controller is amortized by MAC operations in the clusters. In PRU, the expert union unit forms the shared expert set and consumes only 0.2\% of the total power. The remaining routing decision logic, shown as Etc. in Table~\ref{table_4}, selects experts based on the shared expert set and adds only 0.1\% power overhead. In DCU, the compute mask update unit supports computational calibration by updating the masks of loaded channels from 0 to 1. Since this unit consists mainly of simple multiplexers and control logic, it consumes only 0.5\% of the total power. The pruning logic and address generator together consume 4.6\% of the total power, where most of the overhead comes from the pruning logic. The address generator produces addresses for the selected input channels. Although the selected input channels may be fragmented, EdgeXpert fetches weights along the output-channel dimension for each selected input channel, avoiding irregular DRAM accesses. Therefore, EdgeXpert enables memory-efficient LLM inference with minimal overhead.
\section{Discussion}
\label{sec:discussion}

\vspace{-1mm}

\textit{\textbf{Extension to other frameworks and models.}}
EdgeXpert is compatible with diverse speculative decoding frameworks because it optimizes the target model during verification rather than the draft model. It is also applicable to all MoE models, with larger benefits for models with finer-grained experts and for larger models where pruning creates greater channel sparsity.

\textit{\textbf{Edge LLM accelerators.}} Prior edge LLM accelerators primarily focus on model compression through quantization \cite{broca, c-transformer}, pruning \cite{spatten, fact}, and tensor decomposition \cite{broca}. MECLA \cite{mecla} introduces SSMP matrix partition and fine-tuning methods to reduce memory footprint and computation. 
In contrast, EdgeXpert reduces both EMA and computation through EMA-aware expert routing and loading in the MoE and speculative decoding system. 

\section{Conclusion}
\label{sec:conclusion}

\vspace{-1mm}

We propose EdgeXpert, a software-hardware co-designed edge device that addresses the EMA bottleneck caused by the combination of speculative decoding and MoE. We show that naive integration increases expert EMA in both the prefill and decode stages, limiting the benefits of MoE and speculative decoding on edge devices. EdgeXpert addresses this problem with prompt-wise expert reuse in the prefill stage, and depth-aware expert coalescing with computational calibration in the decode stage. Synthesized in Samsung 28nm technology at 800 MHz, EdgeXpert achieves up to 56.3\% latency reduction and 44.1\% energy reduction over prior works \cite{moe-pruner, edgemoe, smolpu}, demonstrating SW-HW co-design for edge devices.




\makeatletter
\renewcommand{\verbatim@font}{\normalfont\ttfamily\fontsize{9}{10}\selectfont}
\makeatother

\appendix

\subsection{Abstract}

We provide the software implementation of the five optimization
techniques in the paper: coarse-grained expert pruning, fine-grained channel pruning,
prompt-wise expert reuse, depth-aware expert coalescing, and computational calibration. All techniques are implemented as PyTorch modules on top of the HuggingFace
implementations of four MoE models (Granite-1B-A400M, OLMoE-1B-7B,
DeepSeek-V2-Lite, and Qwen3-30B-A3B) combined with EAGLE-3 speculative
decoding. A single command-line entry point (\texttt{run.py}) runs each
model and lets evaluators individually enable or disable each technique.

\subsection{Artifact check-list (meta-information)}

{\small
\begin{itemize}
  \item {\bf Algorithm: } Coarse-to-fine pruning, prompt-wise expert
        reuse, depth-aware expert coalescing with computational
        calibration
  \item {\bf Program: } Python
  \item {\bf Model: } Granite-1B-A400M, OLMoE-1B-7B,
        DeepSeek-V2-Lite, Qwen3-30B-A3B, and their EAGLE-3 draft models
  \item {\bf Run-time environment: } Linux, conda, Python 3.13; all
        package versions pinned in \texttt{requirements.txt}
  \item {\bf Hardware: } NVIDIA GPUs (evaluated on RTX A6000 48\,GB);
        1 GPU for Granite/OLMoE, 2 GPUs for DeepSeek/Qwen
  \item {\bf Execution: } Interactive command line; each technique is
        toggled with a \texttt{True}/\texttt{False} flag
  \item {\bf Output: } Generated text printed to stdout
  \item {\bf Experiments: } Run each of the four models with the
        proposed techniques enabled or disabled and a user-defined input
        prompt
  \item {\bf How much disk space required (approximately)?: }
        $\sim$110\,GB (model weights for all four models)
  \item {\bf How much time is needed to prepare workflow
        (approximately)?: } $\sim$10 minutes, excluding model download
  \item {\bf How much time is needed to complete experiments
        (approximately)?: } A few minutes per model after weights are
        downloaded
  \item {\bf Publicly available?: } Yes
  \item {\bf Archived?: }
        \url{https://doi.org/10.5281/zenodo.21481269}
\end{itemize}
}

\subsection{Description}

\subsubsection{How to access}

Access the source code of EdgeXpert's software implementation at:

\begin{itemize}
  \item \url{https://doi.org/10.5281/zenodo.21481269}
\end{itemize}

\subsubsection{Hardware dependencies}

An x86\_64 Linux server with NVIDIA GPUs. Approximate peak VRAM (bf16):
Granite $\sim$3--5\,GB and OLMoE $\sim$16--20\,GB on a single GPU;
DeepSeek-V2-Lite $\sim$34--40\,GB and Qwen3-30B-A3B $\sim$62--70\,GB
across two GPUs.

\subsubsection{Software dependencies}

Linux, conda, and Python 3.13. All package versions are pinned in
\texttt{requirements.txt}.

\subsubsection{Models}

The four MoE base models and their EAGLE-3 draft models are downloaded
automatically from the Hugging Face Hub on first run.

\subsection{Installation}

Follow the setup instructions in the \texttt{README.md} available at the
DOI:

\begin{verbatim}
conda create -n edgexpert python=3.13 -y
conda activate edgexpert
pip install -r requirements.txt
\end{verbatim}

\subsection{Experiment workflow}

Run \texttt{run.py} with the target model and the desired combination of
techniques, e.g.:

\begin{verbatim}
python run.py --model olmoe \
    --coarse_pruning True \
    --fine_pruning True \
    --prompt_wise_expert_reuse True
\end{verbatim}

After the model loads, the program prints \texttt{write prompt:}; type
any prompt to generate text. Per-model example commands are given in the
\texttt{README.md}.

\subsection{Evaluation and expected results}

After loading, the program prints \texttt{model loaded.} and the list of active policies, then enters the prompt loop (\texttt{write prompt:}). For each input prompt, the generated response is printed to stdout. The generated response and its accuracy vary depending on which policies are enabled, as each policy trades output quality for reduced computation. Peak GPU memory can be measured with \texttt{measure\_vram.py} under the same flags.

\subsection{Experiment customization}

Each of the five techniques can be enabled or disabled independently.
Generation parameters (\texttt{-{}-max\_new\_tokens},
\texttt{-{}-temperature}, \texttt{-{}-top\_k}, \texttt{-{}-top\_p},
\texttt{-{}-max\_length}, \texttt{-{}-total\_token}, \texttt{-{}-depth}) and
GPU selection (\texttt{-{}-gpus}) can be overridden on the command line.

\subsection{Notes}

This artifact runs in full precision (bf16). The results reported in the
paper were simulated with 8-bit inputs and 4-bit weights, quantized with
LLM Compressor
(\url{https://github.com/vllm-project/llm-compressor}) using GPTQ with
group size 32.






\bibliographystyle{IEEEtran}
\bibliography{refs}

@article{llama,
  title="{The Llama 3 Herd of Models}",
  author={Grattafiori, Aaron and Dubey, Abhimanyu and Jauhri, Abhinav and Pandey, Abhinav and Kadian, Abhishek and Al-Dahle, Ahmad and Letman, Aiesha and Mathur, Akhil and Schelten, Alan and Vaughan, Alex and others},
  journal={arXiv preprint arXiv:2407.21783},
  year={2024}
}

@article{gpt,
  title="{GPT-4 Technical Report}",
  author={Achiam, Josh and Adler, Steven and Agarwal, Sandhini and Ahmad, Lama and Akkaya, Ilge and Aleman, Florencia Leoni and Almeida, Diogo and Altenschmidt, Janko and Altman, Sam and Anadkat, Shyamal and others},
  journal={arXiv preprint arXiv:2303.08774},
  year={2023}
}

@article{gemini,
  title="{Gemini 1.5: Unlocking multimodal understanding across millions of tokens of context}",
  author={Team, Gemini and Georgiev, Petko and Lei, Ving Ian and Burnell, Ryan and Bai, Libin and Gulati, Anmol and Tanzer, Garrett and Vincent, Damien and Pan, Zhufeng and Wang, Shibo and others},
  journal={arXiv preprint arXiv:2403.05530},
  year={2024}
}

@INPROCEEDINGS{alisa,
  author={Zhao, Youpeng and Wu, Di and Wang, Jun},
  booktitle={2024 ACM/IEEE 51st Annual International Symposium on Computer Architecture (ISCA)}, 
  title="{ALISA: Accelerating Large Language Model Inference via Sparsity-Aware KV Caching}", 
  year={2024},
  volume={},
  number={},
  pages={1005-1017},
  doi={10.1109/ISCA59077.2024.00077}}

@article{squeezed_attention,
  title="{Squeezed attention: Accelerating long context length llm inference}",
  author={Hooper, Coleman and Kim, Sehoon and Mohammadzadeh, Hiva and Maheswaran, Monishwaran and Paik, June and Mahoney, Michael W and Keutzer, Kurt and Gholami, Amir},
  journal={arXiv preprint arXiv:2411.09688},
  year={2024}
}

@inproceedings{mecla,
  title="{MECLA: Memory-Compute-Efficient LLM Accelerator with Scaling Sub-matrix Partition}",
  author={Qin, Yubin and Wang, Yang and Zhao, Zhiren and Yang, Xiaolong and Zhou, Yang and Wei, Shaojun and Hu, Yang and Yin, Shouyi},
  booktitle={2024 ACM/IEEE 51st Annual International Symposium on Computer Architecture (ISCA)},
  pages={1032--1047},
  year={2024},
  organization={IEEE}
}

@article{c-transformer,
  title="{C-Transformer: An Energy-Efficient Homogeneous DNN-Transformer/SNN-Transformer Processor for Large Language Models}",
  author={Kim, Sangyeob and Kim, Sangjin and Jo, Wooyoung and Kim, Soyeon and Hong, Seongyon and Lee, Nayeong and Lee, Jungwan and Yoo, Hoi-Jun},
  journal={IEEE Journal of Solid-State Circuits},
  year={2025},
  publisher={IEEE}
}

@article{deepseekmoe,
  title="{DeepSeekMoE: Towards ultimate expert specialization in mixture-of-experts language models}",
  author={Dai, Damai and Deng, Chengqi and Zhao, Chenggang and Xu, RX and Gao, Huazuo and Chen, Deli and Li, Jiashi and Zeng, Wangding and Yu, Xingkai and Wu, Yu and others},
  journal={arXiv preprint arXiv:2401.06066},
  year={2024}
}

@article{mixtral,
  title="{Mixtral of Experts}",
  author={Jiang, Albert Q and Sablayrolles, Alexandre and Roux, Antoine and Mensch, Arthur and Savary, Blanche and Bamford, Chris and Chaplot, Devendra Singh and Casas, Diego de las and Hanna, Emma Bou and Bressand, Florian and others},
  journal={arXiv preprint arXiv:2401.04088},
  year={2024}
}

@inproceedings{glam,
  title="{GlaM: Efficient scaling of language models with mixture-of-experts}",
  author={Du, Nan and Huang, Yanping and Dai, Andrew M and Tong, Simon and Lepikhin, Dmitry and Xu, Yuanzhong and Krikun, Maxim and Zhou, Yanqi and Yu, Adams Wei and Firat, Orhan and others},
  booktitle={International conference on machine learning},
  pages={5547--5569},
  year={2022},
  organization={PMLR}
}

@inproceedings{llama-moe,
  title="{LLaMA-MoE: Building mixture-of-experts from llama with continual pre-training}",
  author={Zhu, Tong and Qu, Xiaoye and Dong, Daize and Ruan, Jiacheng and Tong, Jingqi and He, Conghui and Cheng, Yu},
  booktitle={Proceedings of the 2024 Conference on Empirical Methods in Natural Language Processing},
  pages={15913--15923},
  year={2024}
}

@article{deepseek-v2,
  title="{Deepseek-v2: A strong, economical, and efficient mixture-of-experts language model}",
  author={Liu, Aixin and Feng, Bei and Wang, Bin and Wang, Bingxuan and Liu, Bo and Zhao, Chenggang and Dengr, Chengqi and Ruan, Chong and Dai, Damai and Guo, Daya and others},
  journal={arXiv preprint arXiv:2405.04434},
  year={2024}
}

@article{olmoe,
  title="{OLMoE: Open mixture-of-experts language models}",
  author={Muennighoff, Niklas and Soldaini, Luca and Groeneveld, Dirk and Lo, Kyle and Morrison, Jacob and Min, Sewon and Shi, Weijia and Walsh, Pete and Tafjord, Oyvind and Lambert, Nathan and others},
  journal={arXiv preprint arXiv:2409.02060},
  year={2024}
}

@article{bild,
  title="{Speculative decoding with big little decoder}",
  author={Kim, Sehoon and Mangalam, Karttikeya and Moon, Suhong and Malik, Jitendra and Mahoney, Michael W and Gholami, Amir and Keutzer, Kurt},
  journal={Advances in Neural Information Processing Systems},
  volume={36},
  pages={39236--39256},
  year={2023}
}

@article{medusa,
  title="{Medusa: Simple llm inference acceleration framework with multiple decoding heads}",
  author={Cai, Tianle and Li, Yuhong and Geng, Zhengyang and Peng, Hongwu and Lee, Jason D and Chen, Deming and Dao, Tri},
  journal={arXiv preprint arXiv:2401.10774},
  year={2024}
}

@article{eagle3,
  title="{EAGLE-3: Scaling up inference acceleration of large language models via training-time test}",
  author={Li, Yuhui and Wei, Fangyun and Zhang, Chao and Zhang, Hongyang},
  journal={arXiv preprint arXiv:2503.01840},
  year={2025}
}

@article{layerskip,
  title="{LayerSkip: Enabling early exit inference and self-speculative decoding}",
  author={Elhoushi, Mostafa and Shrivastava, Akshat and Liskovich, Diana and Hosmer, Basil and Wasti, Bram and Lai, Liangzhen and Mahmoud, Anas and Acun, Bilge and Agarwal, Saurabh and Roman, Ahmed and others},
  journal={arXiv preprint arXiv:2404.16710},
  year={2024}
}

@article{deepseek-r1,
  title="{DeepSeek-R1: Incentivizing reasoning capability in llms via reinforcement learning}",
  author={Guo, Daya and Yang, Dejian and Zhang, Haowei and Song, Junxiao and Zhang, Ruoyu and Xu, Runxin and Zhu, Qihao and Ma, Shirong and Wang, Peiyi and Bi, Xiao and others},
  journal={arXiv preprint arXiv:2501.12948},
  year={2025}
}

@ARTICLE{edgellm,
  author={Xu, Daliang and Yin, Wangsong and Zhang, Hao and Jin, Xin and Zhang, Ying and Wei, Shiyun and Xu, Mengwei and Liu, Xuanzhe},
  journal={IEEE Transactions on Mobile Computing}, 
  title="{EdgeLLM: Fast On-Device LLM Inference With Speculative Decoding}", 
  year={2025},
  volume={24},
  number={4},
  pages={3256-3273},
  doi={10.1109/TMC.2024.3513457}}

@article{eagle1,
  title="{Eagle: Speculative sampling requires rethinking feature uncertainty}",
  author={Li, Yuhui and Wei, Fangyun and Zhang, Chao and Zhang, Hongyang},
  journal={arXiv preprint arXiv:2401.15077},
  year={2024}
}

@article{1st_sd,
  title="{Speculative decoding: Exploiting speculative execution for accelerating seq2seq generation}",
  author={Xia, Heming and Ge, Tao and Wang, Peiyi and Chen, Si-Qing and Wei, Furu and Sui, Zhifang},
  journal={arXiv preprint arXiv:2203.16487},
  year={2022}
}

@inproceedings{2nd_sd,
  title="{Fast inference from transformers via speculative decoding}",
  author={Leviathan, Yaniv and Kalman, Matan and Matias, Yossi},
  booktitle={International Conference on Machine Learning},
  pages={19274--19286},
  year={2023},
  organization={PMLR}
}

@article{3rd_sd,
  title="{Accelerating large language model decoding with speculative sampling}",
  author={Chen, Charlie and Borgeaud, Sebastian and Irving, Geoffrey and Lespiau, Jean-Baptiste and Sifre, Laurent and Jumper, John},
  journal={arXiv preprint arXiv:2302.01318},
  year={2023}
}

@article{sequoia,
  title="{Sequoia: Scalable, robust, and hardware-aware speculative decoding}",
  author={Chen, Zhuoming and May, Avner and Svirschevski, Ruslan and Huang, Yuhsun and Ryabinin, Max and Jia, Zhihao and Chen, Beidi},
  journal={arXiv preprint arXiv:2402.12374},
  year={2024}
}

@inproceedings{specinfer,
author = {Miao, Xupeng and Oliaro, Gabriele and Zhang, Zhihao and Cheng, Xinhao and Wang, Zeyu and Zhang, Zhengxin and Wong, Rae Ying Yee and Zhu, Alan and Yang, Lijie and Shi, Xiaoxiang and Shi, Chunan and Chen, Zhuoming and Arfeen, Daiyaan and Abhyankar, Reyna and Jia, Zhihao},
title = "{SpecInfer: Accelerating Large Language Model Serving with Tree-based Speculative Inference and Verification}",
year = {2024},
isbn = {9798400703867},
publisher = {Association for Computing Machinery},
address = {New York, NY, USA},
url = {https://doi.org/10.1145/3620666.3651335},
doi = {10.1145/3620666.3651335},
pages = {932–949},
numpages = {18},
location = {La Jolla, CA, USA},
series = {ASPLOS '24}
}

@article{moe-pruner,
  title="{MoE-Pruner: Pruning Mixture-of-Experts Large Language Model using the Hints from Its Router}",
  author={Xie, Yanyue and Zhang, Zhi and Zhou, Ding and Xie, Cong and Song, Ziang and Liu, Xin and Wang, Yanzhi and Lin, Xue and Xu, An},
  journal={arXiv preprint arXiv:2410.12013},
  year={2024}
}

@article{moe_alg2,
  title="{Read-ME: Refactorizing LLMs as Router-Decoupled Mixture of Experts with System Co-Design}",
  author={Cai, Ruisi and Ro, Yeonju and Kim, Geon-Woo and Wang, Peihao and Ehteshami Bejnordi, Babak and Akella, Aditya and Wang, Zhangyang and others},
  journal={Advances in Neural Information Processing Systems},
  volume={37},
  pages={116126--116148},
  year={2024}
}

@inproceedings{duplex,
  title="{Duplex: A Device for Large Language Models with Mixture of Experts, Grouped Query Attention, and Continuous Batching}",
  author={Yun, Sungmin and Kyung, Kwanhee and Cho, Juhwan and Choi, Jaewan and Kim, Jongmin and Kim, Byeongho and Lee, Sukhan and Sohn, Kyomin and Ahn, Jung Ho},
  booktitle={2024 57th IEEE/ACM International Symposium on Microarchitecture (MICRO)},
  pages={1429--1443},
  year={2024},
  organization={IEEE}
}

@INPROCEEDINGS{space-mate,
  author={Park, Gwangtae and Song, Seokchan and Sang, Haoyang and Im, Dongseok and Han, Donghyeon and Kim, Sangyeob and Lee, Hongseok and Yoo, Hoi-Jun},
  booktitle={2024 IEEE International Solid-State Circuits Conference (ISSCC)}, 
  title="{20.8 Space-Mate: A 303.5mW Real-Time Sparse Mixture-of-Experts-Based NeRF-SLAM Processor for Mobile Spatial Computing}", 
  year={2024},
  volume={67},
  number={},
  pages={374-376},
  doi={10.1109/ISSCC49657.2024.10454487}}

@INPROCEEDINGS{broca,
  author={Jo, Wooyoung and Hong, Seongyon and Choi, Jiwon and Kwon, Beomseok and Sang, Haoyang and Im, Dongseok and Kim, Sangyeob and Kim, Sangjin and Lee, Taekwon and Yoo, Hoi-Jun},
  booktitle={2025 IEEE International Solid-State Circuits Conference (ISSCC)}, 
  title="{23.7 BROCA: A 52.4-to-559.2mW Mobile Social Agent System-on-Chip with Adaptive Bit-Truncate Unit and Acoustic-Cluster Bit Grouping}", 
  year={2025},
  volume={68},
  number={},
  pages={418-420},
  doi={10.1109/ISSCC49661.2025.10904658}}

@misc{granite,
      title="{Granite 3.0 Language Models}",
      url={https://github.com/ibm-granite/granite-3.0-language-models/},
      author={Granite Team, IBM},
      month={October},
      year={2024}
}

@article{qwen3,
  title="{Qwen3 technical report}",
  author={Yang, An and Li, Anfeng and Yang, Baosong and Zhang, Beichen and Hui, Binyuan and Zheng, Bo and Yu, Bowen and Gao, Chang and Huang, Chengen and Lv, Chenxu and others},
  journal={arXiv preprint arXiv:2505.09388},
  year={2025}
}

@ARTICLE{edgemoe,
  author={Yi, Rongjie and Guo, Liwei and Wei, Shiyun and Zhou, Ao and Wang, Shangguang and Xu, Mengwei},
  journal={IEEE Transactions on Mobile Computing}, 
  title="{EdgeMoE: Empowering Sparse Large Language Models on Mobile Devices}", 
  year={2025},
  volume={24},
  number={8},
  pages={7059-7073},
  doi={10.1109/TMC.2025.3546466}}

@article{specmemo,
  title="{SpecMemo: Speculative Decoding is in Your Pocket}",
  author={Yildirim, Selin and Chen, Deming},
  journal={arXiv preprint arXiv:2506.01986},
  year={2025}
}

@article{moesd,
  title="{MoESD: Unveil Speculative Decoding's Potential for Accelerating Sparse MoE}",
  author={Huang, Zongle and Zhu, Lei and Zhan, Zongyuan and Hu, Ting and Mao, Weikai and Yu, Xianzhi and Liu, Yongpan and Zhang, Tianyu},
  journal={arXiv preprint arXiv:2505.19645},
  year={2025}
}

@inproceedings{asplos_ondevicellm,
  title="{Fast on-device LLM inference with npus}",
  author={Xu, Daliang and Zhang, Hao and Yang, Liming and Liu, Ruiqi and Huang, Gang and Xu, Mengwei and Liu, Xuanzhe},
  booktitle={Proceedings of the 30th ACM International Conference on Architectural Support for Programming Languages and Operating Systems, Volume 1},
  pages={445--462},
  year={2025}
}

@inproceedings{pre-gated_moe,
author = {Hwang, Ranggi and Wei, Jianyu and Cao, Shijie and Hwang, Changho and Tang, Xiaohu and Cao, Ting and Yang, Mao},
title = "{Pre-Gated MoE: An Algorithm-System Co-Design for Fast and Scalable Mixture-of-Expert Inference}",
year = {2025},
isbn = {9798350326581},
publisher = {IEEE Press},
url = {https://doi.org/10.1109/ISCA59077.2024.00078},
doi = {10.1109/ISCA59077.2024.00078},
booktitle = {Proceedings of the 51st Annual International Symposium on Computer Architecture},
pages = {1018–1031},
numpages = {14},
location = {Buenos Aires, Argentina},
series = {ISCA '24}
}

@manual{micron_dram,
title = {{LPDDR4/LPDDR4X SDRAM: MT53E1536M32D4, MT53E768M64D4 Datasheet}},
author = {{Micron Technology, Inc.}},
organization = {{Micron Technology, Inc.}},
year = {2022},
month = jun,
url = {https://www.mouser.com/datasheet/2/671/z3bm_embedded_lpddr4_lpddr4x-3193457.pdf},
note = {Rev. D, datasheet}
}

@article{mt-bench,
  title="{Judging llm-as-a-judge with mt-bench and chatbot arena}",
  author={Zheng, Lianmin and Chiang, Wei-Lin and Sheng, Ying and Zhuang, Siyuan and Wu, Zhanghao and Zhuang, Yonghao and Lin, Zi and Li, Zhuohan and Li, Dacheng and Xing, Eric and others},
  journal={Advances in neural information processing systems},
  volume={36},
  pages={46595--46623},
  year={2023}
}

@article{mmlu,
  title="{Measuring massive multitask language understanding}",
  author={Hendrycks, Dan and Burns, Collin and Basart, Steven and Zou, Andy and Mazeika, Mantas and Song, Dawn and Steinhardt, Jacob},
  journal={arXiv preprint arXiv:2009.03300},
  year={2020}
}

@article{hellaswag,
  title="{Hellaswag: Can a machine really finish your sentence?}",
  author={Zellers, Rowan and Holtzman, Ari and Bisk, Yonatan and Farhadi, Ali and Choi, Yejin},
  journal={arXiv preprint arXiv:1905.07830},
  year={2019}
}

@article{arc_challenge,
  title="{Think you have solved question answering? try arc, the ai2 reasoning challenge}",
  author={Clark, Peter and Cowhey, Isaac and Etzioni, Oren and Khot, Tushar and Sabharwal, Ashish and Schoenick, Carissa and Tafjord, Oyvind},
  journal={arXiv preprint arXiv:1803.05457},
  year={2018}
}

@article{winogrande,
  title={Winogrande: An adversarial winograd schema challenge at scale},
  author={Sakaguchi, Keisuke and Bras, Ronan Le and Bhagavatula, Chandra and Choi, Yejin},
  journal={Communications of the ACM},
  volume={64},
  number={9},
  pages={99--106},
  year={2021},
  publisher={ACM New York, NY, USA}
}

@inproceedings{piqa,
  title={Piqa: Reasoning about physical commonsense in natural language},
  author={Bisk, Yonatan and Zellers, Rowan and Gao, Jianfeng and Choi, Yejin and others},
  booktitle={Proceedings of the AAAI conference on artificial intelligence},
  volume={34},
  number={05},
  pages={7432--7439},
  year={2020}
}

@inproceedings{spatten,
  title={Spatten: Efficient sparse attention architecture with cascade token and head pruning},
  author={Wang, Hanrui and Zhang, Zhekai and Han, Song},
  booktitle={2021 IEEE International Symposium on High-Performance Computer Architecture (HPCA)},
  pages={97--110},
  year={2021},
  organization={IEEE}
}

@inproceedings{fact,
author = {Qin, Yubin and Wang, Yang and Deng, Dazheng and Zhao, Zhiren and Yang, Xiaolong and Liu, Leibo and Wei, Shaojun and Hu, Yang and Yin, Shouyi},
title = "{FACT: FFN-Attention Co-optimized Transformer Architecture with Eager Correlation Prediction}",
year = {2023},
isbn = {9798400700958},
publisher = {Association for Computing Machinery},
address = {New York, NY, USA},
url = {https://doi.org/10.1145/3579371.3589057},
doi = {10.1145/3579371.3589057},
booktitle = {Proceedings of the 50th Annual International Symposium on Computer Architecture},
articleno = {22},
numpages = {14},
location = {Orlando, FL, USA},
series = {ISCA '23}
}

@INPROCEEDINGS{smolpu,
  author={Ha, Sangwoo and Lee, Jingu and Moon, Youngjin and Whang, Sunjoo and Jo, Wooyoung and Park, Gwangtae and Kim, Sangjin and Um, Soyeon and Ryu, Junha and Jo, Yurim and Yoo, Hoi-Jun},
  booktitle={2026 IEEE International Solid-State Circuits Conference (ISSCC)}, 
  title="{SMoLPU: 122.1µJ/Token Sparse MoE-Based Speculative Decoding Language Processing Unit with Adaptive-Offload NPU-CIM Core}", 
  year={2026},
  volume={69},
  number={},
  pages={312-314},
  doi={10.1109/ISSCC49663.2026.11409285}}

@article{quant-based,
  title="{ML-SpecQD: Multi-level speculative decoding with quantized drafts}",
  author={Georganas, Evangelos and Kalamkar, Dhiraj and Kozlov, Alexander and Heinecke, Alexander},
  journal={arXiv preprint arXiv:2503.13565},
  year={2025}
}

@article{c_prun1,
  title="{Specdec++: Boosting speculative decoding via adaptive candidate lengths}",
  author={Huang, Kaixuan and Guo, Xudong and Wang, Mengdi},
  journal={arXiv preprint arXiv:2405.19715},
  year={2024}
}

@article{c_prun2,
  title={Fast best-of-n decoding via speculative rejection},
  author={Sun, Hanshi and Haider, Momin and Zhang, Ruiqi and Yang, Huitao and Qiu, Jiahao and Yin, Ming and Wang, Mengdi and Bartlett, Peter and Zanette, Andrea},
  journal={Advances in Neural Information Processing Systems},
  volume={37},
  pages={32630--32652},
  year={2024}
}

@inproceedings{e_prun1,
  title={Not all experts are equal: Efficient expert pruning and skipping for mixture-of-experts large language models},
  author={Lu, Xudong and Liu, Qi and Xu, Yuhui and Zhou, Aojun and Huang, Siyuan and Zhang, Bo and Yan, Junchi and Li, Hongsheng},
  booktitle={Proceedings of the 62nd Annual Meeting of the Association for Computational Linguistics (Volume 1: Long Papers)},
  pages={6159--6172},
  year={2024}
}

@inproceedings{e_prun2,
  title={MoE-I2: Compressing mixture of experts models through inter-expert pruning and intra-expert low-rank decomposition},
  author={Yang, Cheng and Sui, Yang and Xiao, Jinqi and Huang, Lingyi and Gong, Yu and Duan, Yuanlin and Jia, Wenqi and Yin, Miao and Cheng, Yu and Yuan, Bo},
  booktitle={Findings of the Association for Computational Linguistics: EMNLP 2024},
  pages={10456--10466},
  year={2024}
}

@article{e_prun3,
  title={Efficient expert pruning for sparse mixture-of-experts language models: Enhancing performance and reducing inference costs},
  author={Liu, Enshu and Zhu, Junyi and Lin, Zinan and Ning, Xuefei and Blaschko, Matthew B and Yan, Shengen and Dai, Guohao and Yang, Huazhong and Wang, Yu},
  journal={arXiv preprint arXiv:2407.00945},
  year={2024}
}

@inproceedings{bert,
  title={Bert: Pre-training of deep bidirectional transformers for language understanding},
  author={Devlin, Jacob and Chang, Ming-Wei and Lee, Kenton and Toutanova, Kristina},
  booktitle={Proceedings of the 2019 conference of the North American chapter of the association for computational linguistics: human language technologies, volume 1 (long and short papers)},
  pages={4171--4186},
  year={2019}
}

@inproceedings{vit,
  title={Emerging properties in self-supervised vision transformers},
  author={Caron, Mathilde and Touvron, Hugo and Misra, Ishan and J{\'e}gou, Herv{\'e} and Mairal, Julien and Bojanowski, Piotr and Joulin, Armand},
  booktitle={Proceedings of the IEEE/CVF international conference on computer vision},
  pages={9650--9660},
  year={2021}
}

@inproceedings{rag,
  title={Dense passage retrieval for open-domain question answering},
  author={Karpukhin, Vladimir and Oguz, Barlas and Min, Sewon and Lewis, Patrick and Wu, Ledell and Edunov, Sergey and Chen, Danqi and Yih, Wen-tau},
  booktitle={Proceedings of the 2020 conference on empirical methods in natural language processing (EMNLP)},
  pages={6769--6781},
  year={2020}
}

@misc{minilm,
    title = {all-MiniLM-L6-v2: Sentence Transformers Model},
    author = {Reimers, Nils and Gurevych, Iryna},
    year = {2021},
    howpublished = {\url{https://huggingface.co/sentence-transformers/all-MiniLM-L6-v2}},
    note = {Hugging Face Model Card},
}

@article{cache-prior,
  title={Mixture of cache-conditional experts for efficient mobile device inference},
  author={Skliar, Andrii and van Rozendaal, Ties and Lepert, Romain and Boinovski, Todor and Van Baalen, Mart and Nagel, Markus and Whatmough, Paul and Bejnordi, Babak Ehteshami},
  journal={arXiv preprint arXiv:2412.00099},
  year={2024}
}

@inproceedings{sigma,
  title={Sigma: A sparse and irregular gemm accelerator with flexible interconnects for dnn training},
  author={Qin, Eric and Samajdar, Ananda and Kwon, Hyoukjun and Nadella, Vineet and Srinivasan, Sudarshan and Das, Dipankar and Kaul, Bharat and Krishna, Tushar},
  booktitle={2020 IEEE International Symposium on High Performance Computer Architecture (HPCA)},
  pages={58--70},
  year={2020},
  organization={IEEE}
}

@article{scnn,
  title="{SCNN: An accelerator for compressed-sparse convolutional neural networks}",
  author={Parashar, Angshuman and Rhu, Minsoo and Mukkara, Anurag and Puglielli, Antonio and Venkatesan, Rangharajan and Khailany, Brucek and Emer, Joel and Keckler, Stephen W and Dally, William J},
  journal={ACM SIGARCH computer architecture news},
  volume={45},
  number={2},
  pages={27--40},
  year={2017},
  publisher={ACM New York, NY, USA}
}

@article{gsm8k,
  title={Training verifiers to solve math word problems},
  author={Cobbe, Karl and Kosaraju, Vineet and Bavarian, Mohammad and Chen, Mark and Jun, Heewoo and Kaiser, Lukasz and Plappert, Matthias and Tworek, Jerry and Hilton, Jacob and Nakano, Reiichiro and others},
  journal={arXiv preprint arXiv:2110.14168},
  year={2021}
}

@misc{ttft_tpot,
  author       = {{MLCommons}},
  title        = {{MLPerf Inference interactive benchmark}},
  howpublished = {\url{https://github.com/mlcommons/inference/blob/master/language/llama2-70b/README.md}},
  year         = {2024}
}

@article{moe-spec,
  title="{MoE-Spec: Expert Budgeting for Efficient Speculative Decoding}",
  author={McDanel, Bradley and Li, Steven and Surineni, Sruthikesh and Khaitan, Harshit},
  journal={arXiv preprint arXiv:2602.16052},
  year={2026}
}

@inproceedings{ss-moe,
author = {Zheng, Peirong and Xu, Wenchao and Wang, Haozhao},
title = {Self-Speculative Decoding for On-device MoE Acceleration},
year = {2026},
isbn = {9798400723070},
publisher = {Association for Computing Machinery},
address = {New York, NY, USA},
url = {https://doi.org/10.1145/3774904.3792218},
doi = {10.1145/3774904.3792218},
pages = {5155–5164},
numpages = {10},
location = {United Arab Emirates},
series = {WWW '26}
}

@article{wikitext,
  title="{Pointer Sentinel Mixture Models}",
  author={Merity, Stephen and Xiong, Caiming and Bradbury, James and Socher, Richard},
  journal={arXiv preprint arXiv:1609.07843},
  year={2016}
}

@article{fastervlm,
  title={{[CLS]} Attention is All You Need for Training-Free Visual Token Pruning: Make VLM Inference Faster},
  author={Zhang, Qizhe and Cheng, Aosong and Lu, Ming and Zhuo, Zhiyong and Wang, Minqi and Cao, Jiajun and Guo, Shaobo and She, Qi and Zhang, Shanghang},
  journal={arXiv preprint arXiv:2412.01818v1},
  year={2024},
}

@inproceedings{hiprune,
  title="{HiPrune: Training-Free Visual Token Pruning via Hierarchical Attention in Vision-Language Models (Student Abstract)}",
  author={Liu, Jizhihui and Zhu, Guangdao and Du, Feiyi},
  booktitle={Proceedings of the AAAI Conference on Artificial Intelligence},
  volume={40},
  number={48},
  pages={41275--41277},
  year={2026}
}

@inproceedings{atp-llava,
  title={Atp-llava: Adaptive token pruning for large vision language models},
  author={Ye, Xubing and Gan, Yukang and Ge, Yixiao and Zhang, Xiao-Ping and Tang, Yansong},
  booktitle={Proceedings of the IEEE/CVF Conference on Computer Vision and Pattern Recognition},
  pages={24972--24982},
  year={2025}
}

\end{document}